\documentclass[aps,preprint,groupedaddress,showpacs]{revtex4}
\usepackage{latexsym}

\begin{document}
\bibliographystyle{apsrev}

\title{Vilkovisky-DeWitt Effective Action for Einstein Gravity
on Kaluza-Klein Spacetimes $M^{4}\times S^{N}$}

\author{H.~T.~Cho}
\email{htcho@mail.tku.edu.tw}
\affiliation{
Department of Physics, Tamkang University\\ 
Tamsui, Taipei, Taiwan 251, R.O.C.}

\author{R. Kantowski}
\email{kantowski@mail.nhn.ou.edu}
\affiliation{
Department of Physics and Astronomy, University of Oklahoma\\
Norman, OK 73019, USA}
\date{\today}
\begin{abstract}

We evaluate the divergent part of the Vilkovisky-DeWitt
effective action for Einstein gravity on even-dimensional Kaluza-Klein
spacetimes of the form $M^{4}\times S^{N}$. 
Explicit results are given for $N$=2, 4, and 6.
Trace anomalies for gravitons are
also given for these cases.
Stable  Kaluza-Klein configurations are sought, unsuccessfully,  assuming 
the divergent part of the effective action dominates the dynamics.
\end{abstract}
\maketitle
\newpage

\section{Introduction}

Appelquist and Chodos \cite{AC} were the first to use the effective potential formalism 
to study the  problem of spontaneous
compactification in Kaluza-Klein theories. 
The  hope was that  quantum effects could
explain the smallness of the extra dimensions. It was soon realized that 
results obtained using the standard effective action theory were dependent 
on which quantum gauge fixing condition was used
\cite{RDS,KL}. This non-uniqueness was overcome \cite{BO,HKLT}
by the introduction of a new
effective action formulated by Vilkovisky \cite{GV} and modified by
DeWitt \cite{BD1}. It is now known as the 
Vilkovisky-DeWitt (VD) effective action and has the merit of being 
gauge choice independent. 
Progress in compactification, however,  immediately slowed to a snail's pace 
\cite{BLO,SDOb,BKLM,BKO,BO1,BDO,TV,SDO}, 
because the VD effective
action for gravity involves evaluating determinants of complicated
non-local operators (even at one-loop). In \cite{CK1}, we were able to make 
progress with the 
six-dimensional case of a general background spacetime. We evaluated the 
divergent part
of the VD effective action by extending the four-dimensional
methods of Barvinsky and Vilkovisky \cite{BV}. Due to the complexity of this calculation,
we concluded that it is next to impossible to push this method to higher dimensional
general spacetimes.

In this paper we therefore restrict ourselves to specific
even-dimensional Kaluza-Klein backgrounds of the form
$M^{4}\times S^{N}$. In the next section we briefly review the VD effective action
formalism. We 
then extend the method of Barvinsky and Vilkovisky to these higher dimensional cases and 
expand
the effective action in terms of functional traces of 
various operators. In this way we can extract
the divergent part of the effective action by only considering a 
finite number of terms.

In Section III, the formalism is applied to Einstein gravity. 
The eigenvalues of the operators \cite{RO,CK2}
mentioned above are evaluated for
gravity fields on $M^{4}\times S^{N}$ backgrounds. Using
these eigenvalues, we then extract the divergent parts of the
VD effective action. Because the internal geometry is assumed static the effective action only
gives the effective potential. 
In Section IV two applications are made of these results. First, 
gauge-independent trace anomalies for gravitons are explicitly 
given for $N$=2, 4, and 6. Second, if the divergent part of the effective potential dominates
the dynamics of the internal spheres, self-consistent stable configurations
are shown not to exist. 
Conclusions are given in Section V,
and formulae for the divergent parts
of functional traces, relevant to these calculations,
are listed in the Appendix.

\section{Vilkovisky-DeWitt Effective Action}

In this section we briefly review the formalism of the 
Vilkovisky-DeWitt effective action. 
We follow closely the method
of Barvinsky and Vilkovisky \cite{BV}, as well as 
their notation. 

Consider first a general gauge theory with the action
$S^{G}[\Phi]$, where $\Phi^{i}$ is the set of fields with 
$i$ in the condensed notation representing both the spacetime
and gauge indices. Let $Q^{i}_{\alpha}$ be the generators of
gauge transformations, 
\begin{equation}
\delta\Phi^{i}=Q^{i}_{\alpha}\epsilon^{\alpha},
\end{equation}
where $\epsilon^{\alpha}$ is the gauge parameter. Since the
action $S^{G}$ is gauge invariant,
\begin{equation}
Q^{i}_{\alpha}\frac{\delta S^{G}}{\delta\Phi^{i}}=0.
\label{gauge}
\end{equation}
Up to one-loop, the gauge fixing action in the background field
gauge is given by 
\begin{equation}
S^{GF}=-\frac{1}{2}\chi^{\alpha}c_{\alpha\beta}\chi^{\beta},
\end{equation}
where $\chi^{\alpha}$ is a linear gauge condition, and
$c^{\alpha\beta}$ is a local, invertible matrix. Both 
$\chi^{\alpha}$ and $c_{\alpha\beta}$ may depend on the
background field. The corresponding ghost operator is 
$Q^{i}_{\alpha}(\delta\chi^{\beta}/\delta\Phi^{i})$, and
the one-loop effective action can be written as
\begin{equation}
iW=-\frac{1}{2}{\rm Trln\ }F_{ij}+
{\rm Trln}\left(Q^{i}_{\alpha}
\frac{\delta\chi^{\beta}}{\delta\Phi^{i}}\right),
\label{EA}
\end{equation}
where 
\begin{equation}
F_{ij}=\frac{\delta^{2}S^{G}}{\delta\Phi^{i}\delta\Phi^{j}}-
\frac{\delta\chi^{\alpha}}{\delta\Phi^{i}}c_{\alpha\beta}
\frac{\delta\chi^{\beta}}{\delta\Phi^{j}}.
\end{equation}

However, the one-loop effective action $W$ is gauge dependent 
in general, that is, it depends on the choice of the gauge fixing
action $S^{GF}$ when the background field is not a solution of
the classical equation of motion,
\begin{equation}
{\cal E}_{i}\equiv\frac{\delta S^{G}}{\delta\Phi^{i}}=0.
\end{equation}
This poses a problem using the effective action
formalism in off-shell calculations, for example,
in calculating the trace anomalies for gravitons \cite{CK3}, and in studying
the spontaneous compactification of Kaluza-Klein spaces
\cite{HKLT}. 
This gauge fixing problem can be resolved by using 
the VD effective action, since this effective
action is manifestly independent to the choice 
of gauge conditions. 

At the one-loop level, the VD effective can be obtained simply
by replacing the functional derivative in the ordinary effective
action by a covariant functional derivative
\begin{eqnarray}
\frac{\delta^{2}S^{G}}{\delta\Phi^{i}\delta\Phi^{j}}
&\rightarrow&\frac{D}{\delta\Phi^{i}}
\left(\frac{\delta S^{G}}{\delta\Phi^{j}}\right)\nonumber\\
&=&\frac{\delta^{2}S^{G}}{\delta\Phi^{i}\delta\Phi^{j}}-
\Gamma^{k}_{ij}\frac{\delta S^{G}}{\delta\Phi^{k}},
\end{eqnarray}
where the connection consists of two parts,
\begin{equation}
\Gamma^{k}_{ij}=\left\{
\begin{array}{c}
k \\ i\ j
\end{array}\right\}+T^{k}_{ij}.
\end{equation}
$\left\{ \begin{array}{c} k \\ i\ j \end{array}\right\}$
is the local Christoffel symbol constructed in the usual manner
from a configuration space metric $\gamma_{ij}$, 
\begin{equation}
\left\{
\begin{array}{c}
k \\ i\ j
\end{array}\right\}=\frac{1}{2}\gamma^{kl}
(\gamma_{li,j}+\gamma_{lj,i}-\gamma_{ij,l}),
\end{equation}
where the derivative in 
$\gamma_{li,j}=\delta\gamma_{li}/\delta\Phi^{j}$ represents
the ordinary functional derivative. The configuration  space metric 
is the new ingredient in the VD theory. A prescription for defining
it  has been given by Vilkovisky \cite{GV}. The 
non-local part $T^{k}_{ij}$ of the connection comes from the gauge constraints,
\begin{equation}
T^{k}_{ij}=-2Q^{k}_{\alpha;(i}\gamma_{j)l}N^{\alpha\beta}
Q^{l}_{\beta}+\gamma_{(il}N^{\alpha\mu}Q^{l}_{\mu}Q^{m}_{\alpha}
Q^{k}_{\beta;m}\gamma_{j)n}N^{\beta\nu}Q^{n}_{\nu},
\end{equation}
where the derivative in $Q^{k}_{\beta;m}$ is the covariant 
functional derivative defined with the Christoffel symbol
$\left\{\begin{array}{c}k \\ i\ j\end{array}\right\}$, and
$N^{\alpha\beta}$ is the inverse of $N^{-1}_{\alpha\beta}$,
\begin{equation}
N^{-1}_{\alpha\beta}N^{\beta\gamma}=\delta^{\gamma}_{\alpha},
\end{equation}
with
\begin{equation}
N^{-1}_{\alpha\beta}=\gamma_{ij}Q^{i}_{\alpha}Q^{j}_{\beta}.
\end{equation}
Here we have used the convention of symmetrization such that
$A_{(i}B_{j)}=\frac{1}{2}(A_{i}B_{j}+A_{j}B_{i})$. A detailed
derivation of $T^{k}_{ij}$ can be found, for example, in \cite{GK}. 
Therefore, the one-loop VD effective action can
be written as
\begin{equation}
iW_{unique}=-\frac{1}{2}{\rm Trln}{\cal F}_{ij}+
{\rm Trln}\left(Q^{i}_{\alpha}
\frac{\delta\chi^{\beta}}{\delta\Phi^{i}}\right),
\label{VDEA}
\end{equation}
where
\begin{eqnarray}
{\cal F}_{ij}&=&\frac{D}{\delta\Phi^{i}}
\left(\frac{\delta S^{G}}{\delta\Phi^{j}}\right)-
\frac{\delta\chi^{\alpha}}{\delta\Phi^{i}}c_{\alpha\beta}
\frac{\delta\chi^{\beta}}{\delta\Phi^{j}}\nonumber\\
&=&S^{G}_{,ij}-
\frac{\delta\chi^{\alpha}}{\delta\Phi^{i}}c_{\alpha\beta}
\frac{\delta\chi^{\beta}}{\delta\Phi^{j}}-
\Gamma^{k}_{ij}{\cal E}_{k}.
\end{eqnarray}

To calculate the divergent part of $iW_{unique}$, we separate
the local and the non-local parts of ${\cal F}$. First we 
define the Green's function ${\cal G}$ such that
\begin{equation}
\left\{S^{G}_{,ij}-
\left\{\begin{array}{c}k \\ i\ j\end{array}\right\}
{\cal E}_{k}-
\frac{\delta\chi^{\alpha}}{\delta\Phi^{i}}c_{\alpha\beta}
\frac{\delta\chi^{\beta}}{\delta\Phi^{j}}\right\}
{\cal G}^{jl}=-\delta^{l}_{i}.
\label{defineg}
\end{equation}
Therefore, using $-{\cal G}^{-1}$ to represent the above
operator in the parenthesis,
\begin{eqnarray}
{\cal F}_{ij}&=&-{\cal G}^{-1}_{ij}-T^{k}_{ij}{\cal E}_{k}
\nonumber\\
&=&-{\cal G}^{-1}_{il}(\delta^{l}_{j}+
{\cal G}^{lm}T^{k}_{mj}{\cal E}_{k}).
\end{eqnarray}
The contribution of ${\cal F}$ to the VD effective action
can then be written as
\begin{eqnarray}
-\frac{1}{2}{\rm Trln}{\cal F}&=&
-\frac{1}{2}{\rm Trln}{\cal G}^{-1}-
\frac{1}{2}{\rm Trln}(1+{\cal G}T{\cal E})\nonumber\\
&=&-\frac{1}{2}{\rm Trln}{\cal G}^{-1}-
\frac{1}{2}{\rm Trln}M+\frac{1}{4}{\rm Trln}M^{2}+\cdots,
\label{fexpand}
\end{eqnarray}
where $M^{l}_{j}={\cal G}^{lm}T^{k}_{mj}{\cal E}_{k}$.
The various traces can be evaluated with the help of the 
following identities. From Eq.~(\ref{gauge}),
\begin{eqnarray}
&&\frac{\delta}{\delta\Phi^{j}}
\left(Q^{i}_{\alpha}\frac{\delta S^{G}}{\delta\Phi^{i}}
\right)=0 
\nonumber\\
&\Rightarrow&\left(\frac{\delta Q^{i}_{\alpha}}
{\delta\Phi^{j}}\right){\cal E}_{i}+
Q^{i}_{\alpha}S^{G}_{,ij}=0.
\label{ident}
\end{eqnarray}
From the definition of the operator ${\cal G}^{-1}$ in 
Eq.~(\ref{defineg}),
\begin{equation}
S^{G}_{,ij}=-{\cal G}^{-1}_{ij}+
\left\{ \begin{array}{c} k \\ i\ j \end{array}\right\}
{\cal E}_{k}+
\frac{\delta\chi^{\alpha}}{\delta\Phi^{i}}c_{\alpha\beta}
\frac{\delta\chi^{\beta}}{\delta\Phi^{j}}.
\end{equation}
If we choose the DeWitt background gauge \cite{BD2},
\begin{equation}
\frac{\delta\chi^{\alpha}}{\delta\Phi^{i}}=
-{c^{-1}}^{\alpha\beta}Q^{j}_{\beta}\gamma_{ji},
\end{equation}
we can obtain minimal operators \cite{CK2}
with the appropriate choice
of $c_{\alpha\beta}$. In the DeWitt gauge,
\begin{equation}
\frac{\delta\chi^{\alpha}}{\delta\Phi^{i}}c_{\alpha\beta}
\frac{\delta\chi^{\beta}}{\delta\Phi^{j}}=
\gamma_{ik}Q^{k}_{\alpha}{c^{-1}}^{\alpha\beta}
Q^{l}_{\beta}\gamma_{lj},
\end{equation}
and the ghost operator
\begin{equation}
Q^{i}_{\alpha}\frac{\delta\chi^{\beta}}{\delta\Phi^{i}}=
-N^{-1}_{\alpha\mu}{c^{-1}}^{\mu\beta}.
\label{ghost}
\end{equation}
Therefore, Eq.~(\ref{ident}) becomes
\begin{eqnarray}
&&\left(\frac{\delta Q^{i}_{\alpha}}{\delta\Phi^{j}}\right)
{\cal E}_{i}+
Q^{i}_{\alpha}\left(-{\cal G}^{-1}_{ij}+
\left\{ \begin{array}{c} k \\ i\ j \end{array}\right\}
{\cal E}_{k}+\gamma_{ik}Q^{k}_{\mu}{c^{-1}}^{\mu\nu}
Q^{l}_{\nu}\gamma_{lj}\right)=0 \nonumber\\
&\Rightarrow&
N_{\alpha\beta}{c^{-1}}^{\beta\mu}Q^{i}_{\mu}\gamma_{ij}=
Q^{i}_{\alpha}{\cal G}^{-1}_{ij}-Q^{i}_{\alpha;j}{\cal E}_{i}.
\end{eqnarray}
Multipling both sides by ${\cal G}^{jk}N^{\alpha\nu}$, we have
\begin{equation}
Q^{i}_{\nu}\gamma_{ij}{\cal G}^{jk}=
Q^{k}_{\alpha}N^{\alpha\beta}c_{\beta\nu}-
Q^{i}_{\alpha;j}{\cal E}_{i}{\cal G}^{jk}N^{\alpha\beta}
c_{\beta\nu}.
\label{basic}
\end{equation}
This is the needed basic identity. If we multiply both sides by 
$N^{\mu\gamma}Q^{l}_{\gamma;k}{\cal E}_{l}$,
\begin{eqnarray}
Q^{i}_{\nu}\gamma_{ij}{\cal G}^{jk}Q^{l}_{\gamma;k}
{\cal E}_{l}N^{\mu\gamma}&=&
N^{\mu\gamma}Q^{l}_{\gamma;k}{\cal E}_{l}Q^{k}_{\alpha}
N^{\alpha\beta}c_{\beta\nu}-
N^{\mu\gamma}Q^{i}_{\gamma;k}{\cal E}_{i}{\cal G}^{kl}
Q^{j}_{\alpha;l}{\cal E}_{j}N^{\alpha\beta}c_{\beta\nu}
\nonumber\\
&=&{U_{1}}^{\mu}_{\nu}-{U_{2}}^{\mu}_{\nu},
\label{ident1}
\end{eqnarray}
where
\begin{eqnarray}
{U_{1}}^{\mu}_{\nu}&\equiv&N^{\mu\gamma}Q^{k}_{\gamma}
Q^{l}_{\alpha;k}{\cal E}_{l}N^{\alpha\beta}c_{\beta\nu},\\
{U_{2}}^{\mu}_{\nu}&\equiv&N^{\mu\gamma}Q^{i}_{\gamma;k}
{\cal E}_{i}{\cal G}^{kl}Q^{j}_{\alpha;l}{\cal E}_{j}
N^{\alpha\beta}c_{\beta\nu}.
\end{eqnarray}
We have also used the fact that
\begin{eqnarray}
Q^{i}_{\alpha}Q^{j}_{\gamma;i}{\cal E}_{j}
&=&
Q^{i}_{\alpha}[(Q^{j}_{\gamma}{\cal E}_{j})_{;i}-
Q^{j}_{\gamma}{\cal E}_{j;i}]
\nonumber\\
&=&-Q^{i}_{\alpha}Q^{j}_{\gamma}\left[S^{G}_{,ij}-
\left\{\begin{array}{c}k \\ i\ j\end{array}\right\}
{\cal E}_{k}\right]
\nonumber\\
&=&Q^{i}_{\gamma}Q^{j}_{\alpha,i}{\cal E}_{j},
\end{eqnarray}
is symmetric with respect to the indices $\alpha$ and $\gamma$.
We can obtain yet another identity by multipling the basic
identity, Eq.~(\ref{basic}), with $\gamma_{kl}Q^{l}_{\gamma}$,
\begin{equation}
Q^{i}_{\nu}\gamma_{ij}{\cal G}^{jk}\gamma_{kl}Q^{l}_{\gamma}=
c_{\nu\gamma}-Q^{i}_{\alpha;j}{\cal E}_{i}{\cal G}^{jk}
N^{\alpha\beta}c_{\beta\nu}\gamma_{kl}Q^{l}_{\gamma}.
\label{ident2a}
\end{equation}
Applying again the basic identity, Eq.~(\ref{ident2a}) becomes
\begin{eqnarray}
Q^{i}_{\nu}\gamma_{ik}{\cal G}^{kl}\gamma_{lj}Q^{j}_{\gamma}
&=&c_{\nu\gamma}-Q^{i}_{\alpha;j}{\cal E}_{i}N^{\alpha\beta}
c_{\beta\nu}
(Q^{j}_{\mu}N^{\mu\rho}c_{\rho\gamma}-
Q^{l}_{\mu;k}{\cal E}_{l}{\cal G}^{jk}N^{\mu\rho}c_{\rho\gamma})
\nonumber\\
&=&c_{\nu\beta}(\delta^{\beta}_{\gamma}-{U_{1}}^{\beta}_{\gamma}
+{U_{2}}^{\beta}_{\gamma}).
\label{ident2}
\end{eqnarray}

With these identities we are in a position to evaluate the various
traces in Eq.~(\ref{fexpand}). First,
\begin{eqnarray}
{\rm Tr}M
&=&{\cal G}^{ij}T^{k}_{ij}{\cal E}_{k} 
\nonumber\\
&=&-2({U_{1}}^{\alpha}_{\alpha}-{U_{2}}^{\alpha}_{\alpha})+
{U_{1}}^{\alpha}_{\beta}(\delta^{\beta}_{\alpha}-
{U_{1}}^{\beta}_{\alpha}+{U_{2}}^{\beta}_{\alpha})
\nonumber\\
&=&-{\rm Tr}U_{1}+2{\rm Tr}U_{2}-{\rm Tr}U^{2}_{1}+
{\rm Tr}U_{1}U_{2},
\end{eqnarray}
where we have made use of both the identities in 
Eqs.~(\ref{ident1}) and (\ref{ident2}).
Similarly, we can evaluate the traces of higher powers of $M$.
\begin{eqnarray}
{\rm Tr}M^{2}
&=&-{\rm Tr}U^{2}_{1}+2{\rm Tr}U_{2}+2{\rm Tr}U^{3}_{1}-
2{\rm Tr}U_{1}U_{2}+{\rm Tr}U^{4}_{1}-6{\rm Tr}U^{2}_{1}U_{2}+
4{\rm Tr}U^{2}_{2}
\nonumber\\
&&\ \ -2{\rm Tr}U^{3}_{1}U_{2}+
4{\rm Tr}U_{1}U^{2}_{2}+O({\cal E}^{6}),
\\
{\rm Tr}M^{3}
&=&2{\rm Tr}U^{3}_{1}-3{\rm Tr}U_{1}U_{2}-6{\rm Tr}U^{2}_{1}U_{2}+
6{\rm Tr}U^{2}_{2}
\nonumber\\
&&\ \ -3{\rm Tr}U^{5}_{1}+9{\rm Tr}U^{3}_{1}U_{2}-
3{\rm Tr}U_{1}U^{2}_{2}+O({\cal E}^{6}),
\\
{\rm Tr}M^{4}
&=&-{\rm Tr}U^{4}_{1}+2{\rm Tr}U^{2}_{2}-4{\rm Tr}U^{5}_{1}+
16{\rm Tr}U^{3}_{1}U_{2}-12{\rm Tr}U_{1}U^{2}_{2}+
O({\cal E}^{6}),
\\
{\rm Tr}M^{5}
&=&-{\rm Tr}U^{5}_{1}+5{\rm Tr}U^{3}_{1}U_{2}-
5{\rm Tr}U_{1}U^{2}_{2}+O({\cal E}^{6}).
\end{eqnarray}
Note that we have expanded the various expressions only up to the
fifth power of the first functional derivative ${\cal E}_{i}$ of 
the action. To calculate the divergent part of the VD effective
action, one needs only terms up to some power of ${\cal E}_{i}$,
depending on the dimensionality of the spacetime that one is 
considering.

Finally, the VD effective action in Eq.~(\ref{VDEA})
can be expanded in terms of the 
various traces of the operators $U_{1}$ and $U_{2}$,
\begin{eqnarray}
iW_{unique}&=&-\frac{1}{2}{\rm Trln}{\cal F}+{\rm Trln}N^{-1}
\nonumber\\
&=&-\frac{1}{2}{\rm Trln}{\cal G}^{-1}+{\rm Trln}N^{-1}
\nonumber\\
&&\ \ +\frac{1}{2}{\rm Tr}U_{1}+\frac{1}{4}{\rm Tr}U^{2}_{1}-
\frac{1}{2}{\rm Tr}U_{2}+\frac{1}{6}{\rm Tr}U^{3}_{1}-
\frac{1}{2}{\rm Tr}U_{1}U_{2}+\frac{1}{8}{\rm Tr}U^{4}_{1}
\nonumber\\
&&\ \ -\frac{1}{2}{\rm Tr}U^{2}_{1}U_{2}+
\frac{1}{4}{\rm Tr}U^{2}_{2}+
\frac{1}{10}{\rm Tr}U^{5}_{1}-
\frac{1}{2}{\rm Tr}U^{3}_{1}U_{2}+
\frac{1}{2}{\rm Tr}U_{1}U^{2}_{2}+O({\cal E}^{6}).
\end{eqnarray}
Using this expression 
we evaluate the divergent part
of the VD effective action for Einstein
gravity on Kaluza-Klein spaces $M^{4}\times S^{N}$ 
in the following section.

\section{Einstein gravity on Kaluza-Klein spaces $M^{4}\times S^{N}$}

Here we shall apply the formalism set up in the last section to
the case of Einstein gravity. Because such calculations are very tedious to
carry out for general spacetimes (see \cite{CK1}), we  restrict ourselves to
Kaluza-Klein backgrounds of the form $M^{4}\times S^{N}$. 

\subsection{Einstein gravity}

We start with the n-dimensional Einstein-Hilbert action,
\begin{equation}
S^{G}=\int d^{n}\! x\sqrt{-g}(R-2\Lambda),
\end{equation}
where $\Lambda$ is the cosmological constant. The first functional
derivative
\begin{equation}
{\cal E}^{\mu\nu x}=
\frac{\delta S^{G}}{\delta g_{\mu\nu}(x)}=
-(R^{\mu\nu}-\frac{1}{2}g^{\mu\nu}(R-2\Lambda)).
\end{equation}
The gauge symmetry is the general coordinate invariance,
\begin{equation}
\delta g_{\mu\nu}(x)=\nabla_{\mu}\epsilon_{\nu}+
\nabla_{\nu}\epsilon_{\mu}=
Q_{\mu\nu x, \alpha y}\epsilon^{\alpha}(y),
\end{equation}
where $\epsilon_{\mu}(x)$ is the gauge transformation parameter,  
$\nabla_{\mu}$ is the ordinary covariant derivative, and
\begin{equation}
Q_{\mu\nu x, \alpha y}=-(g_{\mu\alpha}\nabla_{\nu}+
g_{\nu\alpha}\nabla_{\mu})\delta^{n}(x-y).
\end{equation}
Because of this gauge symmetry, the second functional derivative
of the action, $S^{G}_{,ij}$, is a singular operator.
\begin{eqnarray}
&&\frac{\delta S^{G}}{\delta g_{\mu\nu}(x)\delta g_{\alpha\beta}(y)}
\nonumber\\
&=&C^{\mu\nu,\rho\sigma}\left[
\delta^{\alpha\beta}_{\rho\sigma}\Box -
\delta^{(\alpha}_{(\rho}\nabla_{\sigma)}\nabla^{\beta)}-
\delta^{(\alpha}_{(\rho}\nabla^{\beta)}\nabla_{\sigma)}+
g^{\alpha\beta}\nabla_{(\rho}\nabla_{\sigma)}
+2R^{\ (\alpha\ \beta)}_{\rho\ \ \sigma}+
2\delta^{(\alpha}_{(\rho}R^{\beta)}_{\sigma)}\right.
\nonumber\\
&&\ \ \left.
-g^{\alpha\beta}R_{\rho\sigma}-
\frac{2}{n-2}g_{\rho\sigma}R^{\alpha\beta}+
\frac{1}{n-2}g_{\rho\sigma}g^{\alpha\beta}R-
(R-2\Lambda)\delta^{\alpha\beta}_{\rho\sigma}\right]
\delta^{n}(x-y)/\sqrt{-g},
\label{s2}
\end{eqnarray}
where 
\begin{eqnarray}
\delta^{\alpha\beta}_{\rho\sigma}
&=&\delta^{\alpha}_{(\rho}\delta^{\beta}_{\sigma)},
\\
C^{\mu\nu,\alpha\beta}
&=&\frac{1}{4}(g^{\mu\alpha}g^{\nu\beta}+
g^{\mu\beta}g^{\nu\alpha}-g^{\mu\nu}g^{\alpha\beta}).
\end{eqnarray}
To invert (\ref{s2}) we need to pick a gauge, and again as in the last section,
 we choose the DeWitt gauge. The DeWitt gauge requires a metric
for the configuration space of $g_{\mu\nu}(x)$. The one given
by Vilkovisky is \cite{GV}
\begin{equation}
\gamma^{\mu\nu x,\alpha\beta y}=C^{\mu\nu,\alpha\beta}
\sqrt{-g}\ \delta^{n}(x-y),
\end{equation}
in which $C^{\mu\nu,\alpha\beta}$ is exactly the same as the 
factor in front
of the operator $S^{G}_{,ij}$. With this metric $\gamma_{ij}$ and
the choice of $c_{\mu\nu}=\delta_{\mu\nu}$, the DeWitt gauge 
becomes the harmonic gauge and the corresponding graviton operator
in Eq.~(\ref{EA}) is minimal \cite{CK2},
\begin{eqnarray}
F^{\alpha\beta}_{\mu\nu}
&=&\delta^{\alpha\beta}_{\mu\nu}\Box+
2R^{\ (\alpha\ \beta)}_{\mu\ \ \nu}+
2\delta^{(\alpha}_{(\mu}R^{\beta)}_{\nu)}-
g^{\alpha\beta}R_{\mu\nu}
\nonumber\\
&&\ \ -\frac{2}{n-2}g_{\mu\nu}R^{\alpha\beta}+
\frac{1}{n-2}g_{\mu\nu}g^{\alpha\beta}R-
(R-2\Lambda)\delta^{\alpha\beta}_{\mu\nu}.
\end{eqnarray}
Note that
for simplicity we have left out the spacetime coordinate indices,
and we shall do so in the subsequent discussion.
The ghost operator in Eq.~(\ref{ghost}) is now simply,
\begin{equation}
N_{\mu\nu}=-(g_{\mu\nu}\Box+R_{\mu\nu}).
\end{equation}

To find the expression for the local operator ${\cal G}^{-1}$ in
the VD effective action, we need the Christoffel symbol
$\left\{ \begin{array}{c} k \\ i\ j \end{array}\right\}$. From
the metric $\gamma^{\mu\nu,\alpha\beta}$ above,
\begin{equation}
\left\{
\begin{array}{c}
g_{\rho\sigma} \\ g_{\mu\nu}\ g_{\alpha\beta}
\end{array}
\right\}=
\frac{1}{4}g^{\mu\nu}\delta^{\alpha\beta}_{\rho\sigma}+
\frac{1}{4}g^{\alpha\beta}\delta^{\mu\nu}_{\rho\sigma}-
\frac{1}{2}\delta^{\mu(\alpha}_{\rho\sigma}g^{\beta)\nu}-
\frac{1}{2}\delta^{\nu(\alpha}_{\rho\sigma}g^{\beta)\mu}+
\frac{1}{n-2}g_{\rho\sigma}C^{\mu\nu,\alpha\beta}.
\end{equation}
The operator ${\cal G}^{-1}$ has the form
\begin{equation}
{{\cal G}^{-1}}^{\alpha\beta}_{\mu\nu}=
-(\delta^{\alpha\beta}_{\mu\nu}\Box+P^{\alpha\beta}_{\mu\nu}),
\end{equation}
where
\begin{eqnarray}
P^{\alpha\beta}_{\mu\nu}&=&
2R^{\ (\alpha\ \beta)}_{\mu\ \ \nu}-
\frac{1}{2}g^{\alpha\beta}R_{\mu\nu}-
\frac{1}{n-2}g_{\mu\nu}R^{\alpha\beta}
\nonumber\\
&&\ \ +\frac{1}{n-2}(\delta^{\alpha\beta}_{\mu\nu}+
\frac{1}{2}g_{\mu\nu}g^{\alpha\beta})R-  
\frac{n}{2(n-2)}(R-2\Lambda)\delta^{\alpha\beta}_{\mu\nu}.
\end{eqnarray}
Note that we have left out an overall factor 
$C^{\mu\nu,\alpha\beta}$ which is irrelevant in this particular calculation.

Next we would like to find the expressions for the operators
${U_{1}}^{\mu}_{\nu}$ and ${U_{2}}^{\mu}_{\nu}$. To do so we need
to use the following equations.
\begin{eqnarray}
&&(Q_{\alpha\beta,\mu})^{;\rho\sigma}{\cal E}^{\alpha\beta}
\nonumber\\
&=&
\left[
-R^{\nu(\rho}\delta^{\sigma)}_{\mu}\nabla_{\nu}+
R^{(\rho}_{\mu}\nabla^{\sigma)}-
\frac{1}{2}g^{\rho\sigma}R^{\nu}_{\mu}\nabla_{\nu}+
\frac{1}{2}\left(R-\frac{2n}{n-2}\Lambda\right)
\delta^{(\rho}_{\mu}\nabla^{\sigma)}\right.
\nonumber\\
&&\ \ \left.
+\frac{1}{2}(R^{\rho\sigma}-\frac{1}{2}g^{\rho\sigma}R)\nabla_{\mu}+
\frac{n}{2(n-2)}\Lambda g^{\rho\sigma}\nabla_{\mu}+
(\nabla_{\mu}R^{\rho\sigma})-\frac{1}{2}g^{\rho\sigma}(\nabla_{\mu}R)
\right],
\end{eqnarray}
and
\begin{eqnarray}
&&Q_{\rho\sigma,\mu}(Q_{\alpha\beta,\nu})^{;\rho\sigma}
{\cal E}^{\alpha\beta}
\nonumber\\
&=&
g_{\mu\nu}R^{\alpha\beta}\nabla_{\alpha}\nabla_{\beta}-
R_{\mu\nu}\Box-
\frac{1}{2}\left(R-\frac{2n}{n-2}\Lambda\right)
(g_{\mu\nu}\Box+R_{\mu\nu})
\nonumber\\
&&\ \ -R^{\alpha}_{\mu}R_{\alpha\nu}+
R^{\alpha\beta}R_{\mu\alpha\nu\beta}-
(\nabla^{\alpha}R_{\mu\nu})\nabla_{\alpha}+
(\nabla_{\mu}R^{\alpha}_{\nu})\nabla_{\alpha}-
(\nabla_{\nu}R^{\alpha}_{\mu})\nabla_{\alpha}.
\end{eqnarray}
Now, the operator ${U_{1}}^{\mu}_{\nu}$ and ${U_{2}}^{\mu}_{\nu}$
are given by
\begin{eqnarray}
{U_{1}}^{\mu}_{\nu}
&=&N^{\mu}_{\alpha}\left[
\delta^{\alpha}_{\beta}
R^{\sigma\lambda}\nabla_{\sigma}\nabla_{\lambda}-
R^{\alpha}_{\beta}\Box-
\frac{1}{2}\left(R-\frac{2n}{n-2}\Lambda\right)
(\delta^{\alpha}_{\beta}\Box+R^{\alpha}_{\beta})\right.
\nonumber\\
&&\ \ \left.-R^{\alpha}_{\lambda}R^{\lambda}_{\beta}+
R^{\sigma\lambda}R^{\alpha}_{\ \sigma\beta\lambda}-
(\nabla^{\lambda}R^{\alpha}_{\beta})\nabla_{\lambda}+
(\nabla^{\alpha}R^{\lambda}_{\beta})\nabla_{\lambda}-
(\nabla_{\beta}R^{\alpha\lambda})\nabla_{\lambda}\right]
N^{\beta}_{\nu}
\\
{U_{2}}^{\mu}_{\nu}
&=&N^{\nu}_{\alpha}\left\{
\frac{1}{2}\left[
(\nabla^{\alpha}R^{\sigma\lambda})-
(\nabla^{\sigma}R^{\lambda\alpha})-
(\nabla^{\lambda}R^{\sigma\alpha})
\right]+
\left[
D^{\sigma\lambda,\alpha\mu}-
\left(R-\frac{2n}{n-2}\Lambda\right)
C^{\sigma\lambda,\alpha\mu}
\right]\nabla_{\mu}\right\}
\nonumber\\
&&\ \ {\cal G}_{\sigma\lambda,\omega\epsilon}
\left\{
(\nabla_{\beta}R^{\epsilon\omega})-
\frac{1}{2}g^{\epsilon\omega}(\nabla_{\beta}R)-
\left[
D^{\omega\epsilon,\ \gamma}_{\ \ \ \beta}-
\left(R-\frac{2n}{n-2}\Lambda\right)
C^{\omega\epsilon,\ \gamma}_{\ \ \ \beta}
\right]\nabla_{\gamma}
\right\}N^{\beta}_{\mu},
\end{eqnarray}
where $N^{\mu}_{\alpha}$ is the ghost Green's function
\begin{equation}
(\delta^{\mu}_{\alpha}\Box+R^{\mu}_{\alpha})N^{\alpha}_{\nu}
=\delta^{\mu}_{\nu},
\end{equation}
and
\begin{equation}
D^{\sigma\lambda,\alpha\mu}\equiv
\frac{1}{2}(g^{\sigma\lambda}R^{\alpha\mu}-
g^{\alpha\mu}R^{\sigma\lambda}+
g^{\alpha\sigma}R^{\lambda\mu}-
g^{\lambda\mu}R^{\alpha\sigma}+
g^{\alpha\lambda}R^{\sigma\mu}-
g^{\sigma\mu}R^{\alpha\lambda}.
\end{equation}
We have written down the expressions for the operators 
${\cal G}^{-1}$, $N^{-1}$, $U_{1}$, and $U_{2}$,
which are present in the VD effective action. In principle,
the divergent part of the VD effective action can now be 
evaluated by working out the traces of different combinations
of these operators. However, the algebra gets exceedingly tedious
as one goes to higher dimensions. In \cite{CK1}, we have considered
the case of $n=6$, but going to larger values of $n$ seems to be
impossible. In the following analysis, we therefore  restrict
ourselves to the specific cases of Kaluza-Klein spaces of the 
form $M^{4}\times S^{N}$.

\subsection{Eigenvalues of various operators on 
	    $M^{4}\times S^{N}$}

Here we work out the eigenvalues of the operators $N^{-1}$,
${\cal G}^{-1}$, $U_{1}$, and $U_{2}$. To distinguish between
external spacetime and internal space indices, we  use Greek
letters for external spacetime $M^{4}$ and Latin letters for
internal space $S^{N}$. Since the traces of these operators are
divergent in general, a regularization scheme is needed. We 
adopt the method of dimensional regularization by taking the external
spacetime dimension to be a general $d$ and we  take
$d\rightarrow 4-\epsilon$ at the end to extract the divergent
parts. Hence in this subsection we  first assume the spacetime 
dimension to be $M^{d}\times S^{N}$.

First we consider the ghost operator $N^{-1}$: 
\begin{eqnarray}
{N^{-1}}^{\mu}_{\nu}&=&-\delta^{\mu}_{\nu}(\Box_{d}+\Box_{N}),
\\
{N^{-1}}^{a}_{b}&=&-\delta^{a}_{b}[(\Box_{d}+\Box_{N})+R^{a}_{b}]
\nonumber\\
&=&-\delta^{a}_{b}\left[
(\Box_{d}+\Box_{N})+\frac{1}{r^{2}}(N-1)\right],
\end{eqnarray}
where $\Box_{d}$ is the d'Alembertian on $M^{d}$, while 
$\Box_{N}$, $r$, and $R^{a}_{b}$ are the Laplacian, the radius, 
and the Ricci tensor on $S^{N}$, respectively. 
The off-diagonal terms 
${N^{-1}}^{\mu}_{a}$ and ${N^{-1}}^{a}_{\mu}$ vanish. 

The eigenvector $\eta^{\nu}$ of ${N^{-1}}^{\mu}_{\nu}$ is a 
vector with $d$ components on $M^{d}$ and a scalar on $S^{N}$. 
Consequently, \cite{RO}, 
\begin{equation}
{N^{-1}}^{\mu}_{\nu}\eta^{\nu}=(k^{2}-\Lambda_{l})\eta^{\mu},
\label{N1}
\end{equation}
where $k^{\mu}$ is the momentum and 
\begin{equation}
\Lambda_{l}=-\frac{l(l+N-1)}{r^{2}},
\end{equation}
for $l=0,1,2,\cdots$ is the scalar Laplacian eigenvalue on $S^{N}$
with degeneracy
\begin{equation}
D^{(s)}_{l}(N)=\frac{(2l+N-1)(l+N-2)!}{l!(N-1)!}.
\end{equation}
The eigenvectors of ${N^{-1}}^{a}_{b}$ are ${\eta^{b}}^{\rm T}$ and
${\eta^{b}}^{\rm L}$. ${\eta^{b}}^{\rm T}$ is the transverse part of 
$\eta^{b}$, which is a scalar on $M^{d}$ and a vector on $S^{N}$,
with
\begin{equation}
(\nabla_{N})_{b}{\eta^{b}}^{\rm T}=0.
\end{equation}
The eigenvalue of ${\eta^{b}}^{\rm T}$ is 
\begin{equation}
{N^{-1}}^{a}_{b}{\eta^{b}}^{\rm T}=
\left[k^{2}-\Lambda_{l}-\frac{N}{r^{2}}\right]{\eta^{a}}^{\rm T},
\label{N2}
\end{equation}
for $l=1,2,\cdots$ with degeneracy
\begin{equation}
D^{(v)}_{l}(N)=\frac{l(l+N-1)(2l+N-1)(l+N-3)!}{(N-2)!(l+1)!}.
\end{equation}
And the eigenvalue of ${\eta^{b}}^{\rm L}$, the longitudinal part
of $\eta^{b}$, is
\begin{equation}
{N^{-1}}^{a}_{b}{\eta^{b}}^{\rm L}=
\left(k^{2}-\Lambda_{l}-\frac{2}{r^2}(N-1)\right)
{\eta^{a}}^{\rm L},
\label{N3}
\end{equation}
for $l=1,2,\cdots$, with degeneracy $D^{(s)}_{l}(N)$.

For the operator ${\cal G}^{-1}$, the eigenfunctions are
symmetric tensors. First, $\eta_{\alpha\beta}$, which is a 
symmetric tensor on $M^{d}$ and a scalar on $S^{N}$, can
be decomposed \cite{RO} 
into the transverse-traceless (TT) part
$\eta^{\rm TT}_{\alpha\beta}$, the 
longitudinal-transverse-traceless (LTT) part 
$\eta^{\rm LTT}_{\alpha\beta}$, the 
longitudinal-longitudinal-traceless (LLT) part
$\eta^{\rm LLT}_{\alpha\beta}$, and the
trace (Tr) part $\eta^{\rm Tr}_{\alpha\beta}$. 
$\eta^{\rm TT}_{\alpha\beta}$ has $(d-2)(d+1)/2$ components on 
$M^{d}$,
\begin{equation}
{{\cal G}^{-1}}^{\alpha\beta}_{\mu\nu}\eta^{\rm TT}_{\alpha\beta}=
\left(k^{2}-\Lambda_{l}+\frac{1}{2r^{2}}N(N-1)-
\frac{N+d}{N+d-2}\Lambda\right)\eta^{\rm TT}_{\mu\nu}.
\end{equation}
$\eta^{\rm LTT}_{\alpha\beta}$ has $(d-1)$ components
on $M^{d}$,
\begin{equation}
{{\cal G}^{-1}}^{\alpha\beta}_{\mu\nu}\eta^{\rm LTT}_{\alpha\beta}=
\left(k^{2}-\Lambda_{l}+\frac{1}{2r^{2}}(N-1)-
\frac{N+d}{N+d-2}\Lambda\right)\eta^{\rm LTT}_{\mu\nu}.
\end{equation}
$\eta^{\rm LLT}_{\alpha\beta}$ has 1 component on $M^{d}$,
\begin{equation}
{{\cal G}^{-1}}^{\alpha\beta}_{\mu\nu}\eta^{\rm LLT}_{\alpha\beta}=
\left(k^{2}-\Lambda_{l}+\frac{1}{2r^{2}}(N-1)-
\frac{N+d}{N+d-2}\Lambda\right)\eta^{\rm LLT}_{\mu\nu}.
\end{equation}
$\eta^{\rm Tr}_{\alpha\beta}$ also has only 1 component on $M^{d}$,
\begin{equation}
{{\cal G}^{-1}}^{\alpha\beta}_{\mu\nu}\eta^{\rm Tr}_{\alpha\beta}=
\left(k^{2}-\Lambda_{l}+\frac{1}{2r^{2}}
\frac{N(N-1)(N-2)}{N+d-2}-\frac{N+d}{N+d-2}\Lambda\right)
\eta^{\rm Tr}_{\mu\nu}.
\end{equation}
They all have degeneracy $D^{(s)}_{l}(N)$ with $l=0,1,2,\cdots$.

Similarly, $\eta_{ab}$ is a scalar on $M^{d}$ and a symmetric tensor
on $S^{N}$. 
\begin{equation}
{{\cal G}^{-1}}^{cd}_{ab}\eta^{\rm TT}_{cd}=
\left(k^{2}-\Lambda_{l}+\frac{1}{2r^{2}}N(N-1)-
\frac{N+d}{N+d-2}\Lambda\right)\eta^{\rm TT}_{ab},
\end{equation}
for $l=2,3,\cdots$, with degeneracy
\begin{equation}
D^{(t)}_{l}(N)=
\frac{(N+1)(N-2)(l+N)(l-1)(2l+N-1)(l+N-3)!}{2(N-1)!(l+1)!}.
\end{equation}
\begin{equation}
{{\cal G}^{-1}}^{cd}_{ab}\eta^{\rm LTT}_{cd}=
\left(k^{2}-\Lambda_{l}+\frac{1}{2r^{2}}N(N-3)-
\frac{N+d}{N+d-2}\Lambda\right)\eta^{\rm LTT}_{ab},
\end{equation}
for $l=2,3,\cdots$, with degeneracy $D^{(v)}_{l}(N)$.
\begin{equation}
{{\cal G}^{-1}}^{cd}_{ab}\eta^{\rm LLT}_{cd}=
\left(k^{2}-\Lambda_{l}+\frac{1}{2r^{2}}(N-1)(N-4)-
\frac{N+d}{N+d-2}\Lambda\right)\eta^{\rm LLT}_{ab},
\end{equation}
for $l=2,3,\cdots$, with degeneracy $D^{(s)}_{l}(N)$.
\begin{eqnarray}
{{\cal G}^{-1}}^{cd}_{ab}\eta^{\rm Tr}_{cd}&=&
\left(k^{2}-\Lambda_{l}+\frac{1}{2r^{2}}
\frac{N^{3}+N^{2}(2d-7)-2N(3d-7)+4(d-2)}{N+d-2}\right.
\nonumber\\
&&\ \ \left.-\frac{N+d}{N+d-2}\Lambda\right)\eta^{\rm Tr}_{ab},
\end{eqnarray}
for $l=0,1,2,\cdots$, with degeneracy $D^{(s)}_{l}(N)$.
We also have the off-diagonal terms, 
\begin{eqnarray}
{{\cal G}^{-1}}^{ab}_{\mu\nu}\eta^{\rm Tr}_{ab}
&=&-\frac{(N-1)(N-2)d}{2r^{2}(N+d-2)}\eta^{\rm Tr}_{\mu\nu},
\\
{{\cal G}^{-1}}^{\mu\nu}_{ab}\eta^{\rm Tr}_{\mu\nu}
&=&\frac{N(N-1)(d-2)}{2r^{2}(N+d-2)}\eta^{\rm Tr}_{ab}.
\end{eqnarray}
The eigenfunctions $\eta^{\rm Tr}_{\mu\nu}$ and 
$\eta^{\rm Tr}_{ab}$ are seen to be coupled together, and 
${\cal G}^{-1}$ 
forms a $2\times 2$ matrix ${\cal G}^{-1}_{\rm Tr}$
in the subspace of these eigenfunctions. 
The determinant of this matrix is
\begin{eqnarray}
{\rm det}{\cal G}^{-1}_{\rm Tr}
&=&\left[k^{2}-\Lambda_{l}+\frac{1}{2r^{2}}
\frac{N(N-1)(N-2)}{N+d-2}-\frac{N+d}{N+d-2}\Lambda\right]
\nonumber\\
&&\ \ 
\left[k^{2}-\Lambda_{l}+\frac{1}{2r^{2}}
\frac{N^{3}+N^{2}(2d-7)-2N(3d-7)+4(d-2)}{N+d-2}-
\frac{N+d}{N+d-2}\Lambda\right]
\nonumber\\
&&\ \ +\frac{N(N-1)^{2}(N-2)d(d-2)}{4r^{4}(N+d-2)^{2}}.
\label{det}
\end{eqnarray}
There are also eigenfunctions $\eta^{\rm T,T}_{\nu b}$, 
$\eta^{\rm T,L}_{\nu b}$, $\eta^{\rm L,T}_{\nu b}$, and
$\eta^{\rm L,L}_{\nu b}$. $\eta^{\rm T,T}_{\nu b}$ is a transverse
vector on $M^{d}$ as well as on $S^{N}$, and so on. 
$\eta^{\rm T,T}_{\nu b}$ has $(d-1)$ components on $M^{d}$,
\begin{equation}
{{\cal G}^{-1}}^{\nu b}_{\mu a}\eta^{\rm T,T}_{\nu b}=
\frac{1}{2}\left(k^{2}-\Lambda_{l}+\frac{1}{2r^{2}}(N+1)(N-2)-
\frac{N+d}{N+d-2}\Lambda\right)\eta^{\rm T,T}_{\mu a},
\end{equation}
for $l=1,2,\cdots$, with $D^{(v)}_{l}(N)$.
$\eta^{\rm T,L}_{\nu b}$ has $(d-1)$ components on $M^{d}$,
\begin{equation}
{{\cal G}^{-1}}^{\nu b}_{\mu a}\eta^{\rm T,L}_{\nu b}=
\frac{1}{2}\left(k^{2}-\Lambda_{l}+\frac{1}{2r^{2}}(N-1)(N-2)-
\frac{N+d}{N+d-2}\Lambda\right)\eta^{\rm T,L}_{\mu a},
\end{equation}
for $l=1,2,\cdots$, with $D^{(s)}_{l}(N)$.
$\eta^{\rm L,T}_{\nu b}$ has 1 components on $M^{d}$,
\begin{equation}
{{\cal G}^{-1}}^{\nu b}_{\mu a}\eta^{\rm L,T}_{\nu b}=
\frac{1}{2}\left(k^{2}-\Lambda_{l}+\frac{1}{2r^{2}}(N+1)(N-2)-
\frac{N+d}{N+d-2}\Lambda\right)\eta^{\rm L,T}_{\mu a},
\end{equation}
for $l=1,2,\cdots$, with $D^{(v)}_{l}(N)$.
$\eta^{\rm L,L}_{\nu b}$ has 1 components on $M^{d}$,
\begin{equation}
{{\cal G}^{-1}}^{\nu b}_{\mu a}\eta^{\rm L,L}_{\nu b}=
\frac{1}{2}\left(k^{2}-\Lambda_{l}+\frac{1}{2r^{2}}(N-1)(N-2)-
\frac{N+d}{N+d-2}\Lambda\right)\eta^{\rm L,L}_{\mu a},
\end{equation}
for $l=1,2,\cdots$, with $D^{(s)}_{l}(N)$.

For ${U_{1}}^{\mu}_{\nu}$, the eigenfunction $\eta^{\nu}$ is a
vector on $M^{d}$ and a scalar on $S^{N}$. $\eta^{\nu}$ has
$d$ components on $M^{d}$,
\begin{eqnarray}
{U_{1}}^{\mu}_{\nu}\eta^{\nu}&=&
\left[\left(\frac{1}{2r^{2}}N(N-1)-
\frac{N+d}{N+d-2}\Lambda\right)(k^{2})\right.
\nonumber\\
&&\ \ \left.
-\left(\frac{1}{2r^{2}}(N-1)(N-2)-\frac{N+d}{N+d-2}\Lambda\right)
(\Lambda_{l})\right]
\left(\frac{1}{k^{2}-\Lambda_{l}}\right)^{2}\eta^{\mu},
\end{eqnarray}
for $l=0,1,2,\cdots$, with degeneracy $D^{(s)}_{l}(N)$. 

For ${U_{1}}^{a}_{b}$, the eigenfunctions are ${\eta^{b}}^{\rm T}$
and ${\eta^{b}}^{\rm L}$.
\begin{eqnarray}
{U_{1}}^{a}_{b}{\eta^{b}}^{\rm T}&=&
\left[\left(\frac{1}{2r^{2}}(N+2)(N-1)-
\frac{N+d}{N+d-2}\Lambda\right)(k^{2})\right.
\nonumber\\
&&\ \ \left.
-\left(\frac{1}{2r^{2}}N(N-1)-
\frac{N+d}{N+d-2}\Lambda\right)
\left(\Lambda_{l}+\frac{N}{r^{2}}\right)\right]
\left(\frac{1}{k^{2}-\Lambda_{l}-\frac{N}{r^{2}}}\right)^{2}
{\eta^{a}}^{\rm T},
\end{eqnarray}
for $l=1,2,\cdots$, with degeneracy $D^{(v)}_{l}(N)$. And
\begin{eqnarray}
{U_{1}}^{a}_{b}{\eta^{b}}^{\rm L}&=&
\left[\left(\frac{1}{2r^{2}}(N+2)(N-1)-
\frac{N+d}{N+d-2}\Lambda\right)(k^{2})\right.
\nonumber\\
&&\ \ \left.
-\left(\frac{1}{2r^{2}}N(N-1)-
\frac{N+d}{N+d-2}\Lambda\right)
\left(\Lambda_{l}+\frac{2}{r^{2}}(N-1)\right)\right]
\nonumber\\
&&\ \ \ \ 
\left(\frac{1}{k^{2}-\Lambda_{l}-\frac{2}{r^{2}}(N-1)}\right)^{2}
{\eta^{a}}^{\rm L},
\end{eqnarray}
for $l=1,2,\cdots$, with degeneracy $D^{(s)}_{l}(N)$.        

The eigenvalues of $U_{2}$ are more complicated. After lengthy
calculations, we obtained the following results. For the eigenfunction
${\eta^{\nu}}^{\rm T}$, which has $(d-1)$ components on $M^{d}$,
\begin{eqnarray}
&&{U_{2}}^{\mu}_{\nu}{\eta^{\nu}}^{\rm T}
\nonumber\\
&=&
\left[\left(\frac{1}{2r^{2}}N(N-1)-
\frac{N+d}{N+d-2}\Lambda\right)^{2}(k^{2})
\left(\frac{1}{k^{2}-\Lambda_{l}+\frac{1}{2r^{2}}N(N-1)-
\frac{N+d}{N+d-2}\Lambda}\right)
\right.\nonumber\\
&&\ \ -\left(\frac{1}{2r^{2}}(N-1)(N-2)-
\frac{N+d}{N+d-2}\Lambda\right)^{2}(\Lambda_{l})
\nonumber\\
&&\ \ \ \ \left.
\left(\frac{1}{k^{2}-\Lambda_{l}+\frac{1}{2r^{2}}(N-1)(N-2)-
\frac{N+d}{N+d-2}\Lambda}\right)\right]
\left(\frac{1}{k^{2}-\Lambda_{l}}\right)^{2}{\eta^{\mu}}^{\rm T},
\end{eqnarray}
for $l=0,1,2,\cdots$, with degeneracy $D^{(s)}_{l}(N)$. 
For eigenfunction ${\eta^{b}}^{\rm T}$, which is a scalar on
$M^{d}$,
\begin{eqnarray}
&&{U_{2}}^{a}_{b}{\eta^{b}}^{\rm T}
\nonumber\\
&=&
\left[\left(\frac{1}{2r^{2}}(N+2)(N-1)-
\frac{N+d}{N+d-2}\Lambda\right)^{2}(k^{2})
\left(\frac{1}{k^{2}-\Lambda_{l}+\frac{1}{2r^{2}}(N+1)(N-2)-
\frac{N+d}{N+d-2}\Lambda}\right)
\right.\nonumber\\
&&\ \ -\left(\frac{1}{2r^{2}}N(N-1)-
\frac{N+d}{N+d-2}\Lambda\right)^{2}
\left(\Lambda_{l}+\frac{N}{r^{2}}\right)
\nonumber\\
&&\ \ \ \ \left.
\left(\frac{1}{k^{2}-\Lambda_{l}+\frac{1}{2r^{2}}N(N-3)-
\frac{N+d}{N+d-2}\Lambda}\right)\right]
\left(\frac{1}{k^{2}-\Lambda_{l}-\frac{N}{r^{2}}}\right)^{2}
{\eta^{a}}^{\rm T},
\end{eqnarray}
for $l=1,2,\cdots$, with degeneracy $D^{(v)}_{l}(N)$.        

For the eigenfunctions ${\eta^{\nu}}^{\rm L}$ and                                           
${\eta^{b}}^{\rm L}$, they are coupled together. For $l=0$,
\begin{eqnarray}
&&{U_{2}}^{\mu}_{\nu}{\eta^{\nu}}^{\rm L}
\nonumber\\
&=&
\left[2\left(\frac{d-1}{d}\right)
\left(\frac{1}{2r^{2}}N(N-1)-\frac{N+d}{N+d-2}\right)^{2}
\left(\frac{1}{k^{2}}\right)
\left(\frac{1}{k^{2}+\frac{1}{2r^{2}}N(N-1)-
\frac{N+d}{N+d-2}\Lambda}\right)\right.
\nonumber\\
&&\ \ -\left(\frac{d-2}{d}\right)
\left(\frac{1}{2r^{2}}N(N-1)-\frac{N+d}{N+d-2}\Lambda\right)
\left(\frac{1}{2r^{2}}\frac{N(N-1)(N-2)}{N+d-2}-
\frac{N+d}{N+d-2}\Lambda\right) 
\nonumber\\
&&\ \ \ \left(k^{2}+\frac{1}{2r^{2}}
\frac{N^{3}+N^{2}(2d-7)-2N(3d-7)+4(d-2)}{N+d-2}-
\frac{N+d}{N+d-2}\Lambda\right)
\left(\frac{1}{k^{2}}\right)\frac{1}{{\rm det}{\cal G}^{-1}_{l=0}}
\nonumber\\
&&\ \ -\frac{N(N-1)^{2}(N-2)(d-2)^{2}}{4r^{4}(N+d-2)^{2}}
\left(\frac{1}{2r^{2}}N(N-1)-
\frac{N+d}{N+d-2}\Lambda\right)
\left(\frac{1}{k^{2}}\right)\frac{1}
{({\rm det}{\cal G}^{-1}_{\rm Tr})_{l=0}}
\nonumber\\
&&\ \ \left.-\frac{N(N-1)(d-2)}{2r^{2}(N+d-2)}
\left(\frac{1}{2r^{2}}(N-1)(N-2)-
\frac{N+d}{N+d-2}\Lambda\right)
\frac{1}{({\rm det}{\cal G}^{-1}_{\rm Tr})_{l=0}}
\right]{\eta^{\mu}}^{\rm L},
\nonumber\\
\end{eqnarray}
where $({\rm det}{\cal G}^{-1}_{\rm Tr})_{l=0}$ is the 
expression ${\rm det}{\cal G}^{-1}_{\rm Tr}$ in 
Eq.~(\ref{det}) evaluated 
at $l=0$. The other eigenfunctions do not contribute in this
case. For $l=1,2,\cdots$,
\begin{eqnarray}
&&{U_{2}}^{\mu}_{\nu}{\eta^{\nu}}^{\rm L}
\nonumber\\
&=&
\left[2\left(\frac{d-1}{d}\right)
\left(\frac{1}{2r^{2}}N(N-1)-\frac{N+d}{N+d-2}\right)^{2}
(k^{2})\right.
\nonumber\\
&&\ \ \ \ \left(\frac{1}{k^{2}-\Lambda_{l}}\right)^{2}
\left(\frac{1}{k^{2}-\Lambda_{l}+\frac{1}{2r^{2}}N(N-1)-
\frac{N+d}{N+d-2}\Lambda}\right)
\nonumber\\
&&\ \ -\left(\frac{1}{2r^{2}}(N-1)(N-2)-
\frac{N+d}{N+d-2}\Lambda\right)^{2}(\Lambda_{l})
\nonumber\\
&&\ \ \ \ 
\left(\frac{1}{k^{2}-\Lambda_{l}}\right)^{2}
\left(\frac{1}{k^{2}-\Lambda_{l}+\frac{1}{2r^{2}}(N-1)(N-2)-
\frac{N+d}{N+d-2}\Lambda}\right)
\nonumber\\
&&\ \ -\left(\frac{d-2}{d}\right)
\left(\frac{1}{2r^{2}}N(N-1)-\frac{N+d}{N+d-2}\Lambda\right)
\left(\frac{1}{2r^{2}}\frac{N(N-1)(N-2)}{N+d-2}-
\frac{N+d}{N+d-2}\Lambda\right)(k^{2}) 
\nonumber\\
&&\ \ \ \left(k^{2}-\Lambda_{l}+\frac{1}{2r^{2}}
\frac{N^{3}+N^{2}(2d-7)-2N(3d-7)+4(d-2)}{N+d-2}-
\frac{N+d}{N+d-2}\Lambda\right)
\nonumber\\
&&\ \ \ \ 
\left(\frac{1}{k^{2}-\Lambda_{l}}\right)^{2}
\frac{1}{{\rm det}{\cal G}^{-1}_{\rm Tr}}
\nonumber\\
&&\ \ -\frac{N(N-1)^{2}(N-2)(d-2)^{2}}{4r^{4}(N+d-2)^{2}}
\left(\frac{1}{2r^{2}}N(N-1)-
\frac{N+d}{N+d-2}\Lambda\right)(k^{2})
\left(\frac{1}{k^{2}-\Lambda_{l}}\right)^{2}
\frac{1}{{\rm det}{\cal G}^{-1}_{\rm Tr}}
\nonumber\\
&&\ \ \left.-\frac{N(N-1)(d-2)}{2r^{2}(N+d-2)}
\left(\frac{1}{2r^{2}}(N-1)(N-2)-
\frac{N+d}{N+d-2}\Lambda\right)(k^{2})
\left(\frac{1}{k^{2}-\Lambda_{l}}\right)
\frac{1}{{\rm det}{\cal G}^{-1}_{\rm Tr}}
\right]{\eta^{\mu}}^{\rm L},
\nonumber\\
\\
&&{U_{2}}^{a}_{\nu}{\eta^{\nu}}^{\rm L}
\nonumber\\
&=&
\left[\left(\frac{1}{2r^{2}}(N+2)(N-1)-
\frac{N+d}{N+d-2}\Lambda\right)
\left(\frac{1}{2r^{2}}(N-1)(N-2)-
\frac{N+d}{N+d-2}\Lambda\right)(k^{2})\right.
\nonumber\\
&&\ \ \ \ 
\left(\frac{1}{k^{2}-\Lambda_{l}}\right)
\left(\frac{1}{k^{2}-\Lambda_{l}-
\frac{2}{r^{2}}(N-1)}\right)
\left(\frac{1}{k^{2}-\Lambda_{l}+
\frac{1}{2r^{2}}(N-1)(N-2)-
\frac{N+d}{N+d-2}\Lambda}\right)
\nonumber\\
&&\ \ -\left(\frac{1}{2r^{2}}(N+2)(N-1)-
\frac{N+d}{N+d-2}\Lambda\right)
\left(\frac{1}{2r^{2}}\frac{N(N-1)(N-2)}{N+d-2}-
\frac{N+d}{N+d-2}\Lambda\right)(k^{2}) 
\nonumber\\
&&\ \ \ \left(k^{2}-\Lambda_{l}+\frac{1}{2r^{2}}
\frac{N^{3}+N^{2}(2d-7)-2N(3d-7)+4(d-2)}{N+d-2}-
\frac{N+d}{N+d-2}\Lambda\right)
\nonumber\\
&&\ \ \ \ 
\left(\frac{1}{k^{2}-\Lambda_{l}}\right)
\left(\frac{1}{k^{2}-\Lambda_{l}-\frac{2}{r^{2}}(N-1)}\right) 
\frac{1}{{\rm det}{\cal G}^{-1}_{\rm Tr}}
\nonumber\\
&&\ \ -\frac{N(N-1)^{2}(N-2)d(d-2)}{4r^{4}(N+d-2)^{2}}
\left(\frac{1}{2r^{2}}(N+2)(N-1)-
\frac{N+d}{N+d-2}\Lambda\right)(k^{2})
\nonumber\\
&&\ \ \ \ 
\left(\frac{1}{k^{2}-\Lambda_{l}}\right)
\left(\frac{1}{k^{2}-\Lambda_{l}-\frac{2}{r^{2}}(N-1)}\right) 
\frac{1}{{\rm det}{\cal G}^{-1}_{\rm Tr}}
\nonumber\\
&&\ \ -\frac{(N-1)(N-2)(d-2)}{2r^{2}(N+d-2)}
\left(\frac{1}{2r^{2}}N(N-1)-
\frac{N+d}{N+d-2}\Lambda\right)(k^{2})
\nonumber\\
&&\ \ \ \ \left.
\left(\frac{1}{k^{2}-\Lambda_{l}-\frac{2}{r^{2}}(N-1)}\right)
\frac{1}{{\rm det}{\cal G}^{-1}_{\rm Tr}}
\right]{\eta^{a}}^{\rm L},
\nonumber\\
\\
&&{U_{2}}^{\mu}_{b}{\eta^{b}}^{\rm L}
\nonumber\\
&=&
\left[-\left(\frac{1}{2r^{2}}(N-1)(N-2)-
\frac{N+d}{N+d-2}\Lambda\right)
\left(\frac{1}{2r^{2}}(N+2)(N-1)-
\frac{N+d}{N+d-2}\Lambda\right)(\Lambda_{l})\right.
\nonumber\\
&&\ \ \ \ 
\left(\frac{1}{k^{2}-\Lambda_{l}}\right)
\left(\frac{1}{k^{2}-\Lambda_{l}-
\frac{2}{r^{2}}(N-1)}\right)
\left(\frac{1}{k^{2}-\Lambda_{l}+
\frac{1}{2r^{2}}(N-1)(N-2)-
\frac{N+d}{N+d-2}\Lambda}\right)
\nonumber\\
&&\ \ +\left(\frac{1}{2r^{2}}(N-1)(N-2)-
\frac{N+d}{N+d-2}\Lambda\right)
\nonumber\\
&&\ \ \ \ 
\left(\frac{1}{2r^{2}}\frac{N(N-1)(N+2d-2)}{N+d-2}-
\frac{N+d}{N+d-2}\Lambda\right)(\Lambda_{l}) 
\nonumber\\
&&\ \ \ \left(k^{2}-\Lambda_{l}+\frac{1}{2r^{2}}
\frac{N(N-1)(N-2)}{N+d-2}-
\frac{N+d}{N+d-2}\Lambda\right)
\left(\frac{1}{k^{2}-\Lambda_{l}}\right)
\nonumber\\
&&\ \ \ \ 
\left(\frac{1}{k^{2}-\Lambda_{l}-\frac{2}{r^{2}}(N-1)}\right) 
\frac{1}{{\rm det}{\cal G}^{-1}_{\rm Tr}}
\nonumber\\
&&\ \ +\frac{N(N-1)^{2}(N-2)d(d-2)}{4r^{4}(N+d-2)^{2}}
\left(\frac{1}{2r^{2}}(N-1)(N-2)-
\frac{N+d}{N+d-2}\Lambda\right)(\Lambda_{l})
\nonumber\\
&&\ \ \ \ 
\left(\frac{1}{k^{2}-\Lambda_{l}}\right)
\left(\frac{1}{k^{2}-\Lambda_{l}-\frac{2}{r^{2}}(N-1)}\right) 
\frac{1}{{\rm det}{\cal G}^{-1}_{\rm Tr}}
\nonumber\\
&&\ \ -\frac{(N-1)(N-2)(d-2)}{2r^{2}(N+d-2)}
\left(\frac{1}{2r^{2}}N(N-1)-
\frac{N+d}{N+d-2}\Lambda\right)(\Lambda_{l})
\nonumber\\
&&\ \ \ \left(k^{2}-\Lambda_{l}-\frac{2}{r^{2}}
\frac{N^{2}+N(d-3)-(d-2)}{N+d-2}\right)
\left(\frac{1}{k^{2}-\Lambda_{l}}\right)
\nonumber\\
&&\ \ \ \ \left.
\left(\frac{1}{k^{2}-\Lambda_{l}-\frac{2}{r^{2}}(N-1)}\right) 
\frac{1}{{\rm det}{\cal G}^{-1}_{\rm Tr}}
\right]{\eta^{\mu}}^{\rm L},
\nonumber\\
\\
&&{U_{2}}^{a}_{b}{\eta^{b}}^{\rm L}
\nonumber\\
&=&
\left[-2\left(\frac{N-1}{N}\right)
\left(\frac{1}{2r^{2}}N(N-1)-\frac{N+d}{N+d-2}\right)^{2}
\left(\Lambda_{l}+\frac{N}{r^{2}}\right)\right.
\nonumber\\
&&\ \ \ \ 
\left(\frac{1}{k^{2}-\Lambda_{l}-\frac{2}{r^{2}}(N-1)}\right)^{2}
\left(\frac{1}{k^{2}-\Lambda_{l}+\frac{1}{2r^{2}}(N-1)(N-4)-
\frac{N+d}{N+d-2}\Lambda}\right)
\nonumber\\
&&\ \ \left.+(\frac{1}{2r^{2}}(N+2)(N-1)-
\frac{N+d}{N+d-2}\Lambda\right)^{2}(k^{2})
\nonumber\\
&&\ \ \ \
\left(\frac{1}{k^{2}-\Lambda_{l}-\frac{2}{r^{2}}(N-1)}\right)^{2}
\left(\frac{1}{k^{2}-\Lambda_{l}+\frac{1}{2r^{2}}(N-1)(N-2)-
\frac{N+d}{N+d-2}\Lambda}\right)
\nonumber\\
&&\ \ +\left(\frac{N-2}{N}\right)
\left(\frac{1}{2r^{2}}N(N-1)-\frac{N+d}{N+d-2}\Lambda\right)
\nonumber\\
&&\ \ \ \
\left(\frac{1}{2r^{2}}\frac{N(N-1)(N+2d-2)}{N+d-2}-
\frac{N+d}{N+d-2}\Lambda\right)(\Lambda_{l}) 
\nonumber\\
&&\ \ \ \left(k^{2}-\Lambda_{l}+\frac{1}{2r^{2}}
\frac{N(N-1)(N-2)}{N+d-2}-
\frac{N+d}{N+d-2}\Lambda\right)
\left(\frac{1}{k^{2}-\Lambda_{l}-\frac{2}{r^{2}}(N-1)}\right)^{2}
\frac{1}{{\rm det}{\cal G}^{-1}_{\rm Tr}}
\nonumber\\
&&\ \ +\frac{(N-1)^{2}(N-2)^{2}d(d-2)}{4r^{4}(N+d-2)^{2}}
\left(\frac{1}{2r^{2}}N(N-1)-
\frac{N+d}{N+d-2}\Lambda\right)(\Lambda_{l})
\nonumber\\
&&\ \ \ \
\left(\frac{1}{k^{2}-\Lambda_{l}-\frac{2}{r^{2}}(N-1)}\right)^{2}
\frac{1}{{\rm det}{\cal G}^{-1}_{\rm Tr}}
\nonumber\\
&&\ \ -\frac{(N-1)(N-2)d}{2r^{2}(N+d-2)}
\left(\frac{1}{2r^{2}}(N+2)(N-1)-
\frac{N+d}{N+d-2}\Lambda\right)(\Lambda_{l})
\nonumber\\
&&\ \ \ \ \left.
\left(k^{2}-\Lambda_{l}-
\frac{2}{r^{2}}\frac{N^{2}+N(d-3)-(d-2)}{N+d-2}\right)
\left(\frac{1}{k^{2}-\Lambda_{l}-\frac{2}{r^{2}}(N-1)}\right)^{2}
\frac{1}{{\rm det}{\cal G}^{-1}_{\rm Tr}}
\right]{\eta^{b}}^{\rm L}.
\nonumber\\
\end{eqnarray}
Both ${\eta^{\nu}}^{\rm L}$ and ${\eta^{b}}^{\rm L}$ have 
degeneracy $D^{(s)}_{l}(N)$.

\subsection{VD effective actions for
	    gravitons on $M^{4}\times S^{N}$}

With the eigenvalues of the various operators in the last
subsection and the divergent parts of their traces listed
in the Appendix, we can evaluate the divergent parts of the
VD effective action on $M^{4}\times S^{N}$. 

Let us start with Trln$N^{-1}$, the ghost contribution to the 
VD effection action. From Eqs.~(\ref{N1}), (\ref{N2}), and 
(\ref{N3}), we obtain on
$M^{4}\times S^{N}$,
\begin{eqnarray}
\left.{\rm Trln}N^{-1}\right\vert^{div}
&=&d\sum_{k}\sum^{\infty}_{l=0}D^{(s)}_{l}(N)
{\rm ln}(k^{2}-\Lambda_{l})+
\sum_{k}\sum^{\infty}_{l=1}D^{(v)}_{l}(N)
{\rm ln}\left(k^{2}-\Lambda_{l}-\frac{N}{r^{2}}\right)
\nonumber\\
&&\ \ \left.+\sum_{k}\sum^{\infty}_{l=1}D^{(s)}_{l}(N)
{\rm ln}\left(k^{2}-\Lambda_{l}-\frac{2}{r^{2}}(N-1)\right)
\right\vert^{div}.
\end{eqnarray}
If we start the summations over $l$ from $l=0$, 
\begin{eqnarray}
\left.{\rm Trln}N^{-1}\right\vert^{div}
&=&d\sum_{k}\sum^{\infty}_{l=0}D^{(s)}_{l}(N)
{\rm ln}(k^{2}-\Lambda_{l})+
\sum_{k}\sum^{\infty}_{l=0}D^{(v)}_{l}(N)
{\rm ln}\left(k^{2}-\Lambda_{l}-\frac{N}{r^{2}}\right)
\nonumber\\
&&\ \ +\sum_{k}\sum^{\infty}_{l=0}D^{(s)}_{l}(N)
{\rm ln}\left(k^{2}-\Lambda_{l}-\frac{2}{r^{2}}(N-1)\right)
\nonumber\\
&&\left.-\delta_{N2}\sum_{k}
{\rm ln}\left(k^{2}-\frac{N}{r^{2}}\right)-
\sum_{k}{\rm ln}\left(k^{2}-\frac{2}{r^{2}}(N-1)\right)
\right\vert^{div}
\nonumber\\
&=&d\sum_{k}\sum^{\infty}_{l=0}D^{(s)}_{l}(N)
{\rm ln}(k^{2}-\Lambda_{l})
\nonumber\\
&&\ \ +\sum_{k}\sum^{\infty}_{l=0}D^{(v)}_{l}(N)
{\rm ln}(k^{2}-\Lambda_{l})
-\sum_{k}\sum^{\infty}_{l=0}D^{(v)}_{l}(N)
\sum^{\infty}_{q=1}\frac{1}{q}\left(\frac{N}{r^{2}}\right)^{q}
\left(\frac{1}{k^{2}-\Lambda_{l}}\right)^{q}
\nonumber\\
&&+\sum_{k}\sum^{\infty}_{l=0}D^{(s)}_{l}(N)
{\rm ln}(k^{2}-\Lambda_{l})-
\sum_{k}\sum^{\infty}_{l=0}D^{(s)}_{l}(N)
\sum^{\infty}_{q=1}\frac{1}{q}
\left(\frac{2(N-1)}{r^{2}}\right)^{q}
\left(\frac{1}{k^{2}-\Lambda_{l}}\right)^{q}
\nonumber\\
&&-\delta_{N2}\sum_{k}{\rm ln}k^{2}+
\delta_{N2}\sum_{k}\sum^{\infty}_{q=1}\frac{1}{q}
\left(\frac{N}{r^{2}}\right)^{q}
\left(\frac{1}{k^{2}}\right)^{q}
\nonumber\\
&&-\sum_{k}{\rm ln}k^{2}+\left.\sum_{k}\sum^{\infty}_{q=1}
\frac{1}{q}\left(\frac{2(N-1)}{r^{2}}\right)^{q}
\left(\frac{1}{k^{2}}\right)^{q}\right\vert^{div},
\end{eqnarray}
where we have Taylor-expanded the logarithmic function. Using
the results in the Appendix,
\begin{eqnarray}
&&\left.{\rm Trln}N^{-1}\right\vert^{div}
\nonumber\\
&=&
\frac{iV_{4}}{(4\pi r^{2})^{2}\epsilon}
\left[F^{(v)}(N)+5F^{(s)}(N)-
\sum^{\infty}_{q=1}\frac{1}{q}
\left(\frac{N}{r^{2}}\right)^{q}G^{(v)}(0,q,N)\right.
\nonumber\\
&&\ \ \left.
-\sum^{\infty}_{q=1}\frac{1}{q}
\left(\frac{2(N-1)}{r^{2}}\right)^{q}G^{(s)}(0,q,N)
+\delta_{N2}\left(\frac{1}{2}\right)
\left(\frac{N}{r^{2}}\right)^{2}(2r^{4})+
\left(\frac{1}{2}\right)\left(\frac{2(N-1)}{r^{2}}\right)^{2}
(2r^{4})\right].
\nonumber\\
\end{eqnarray}
Therefore, for $N=2$,
\begin{equation}
\left.{\rm Trln}N^{-1}\right\vert^{div}=
\frac{iV_{4}}{(4\pi r^{2})^{2}\epsilon}\left[-\frac{24}{35}\right].
\end{equation}
For $N=4$,
\begin{equation}
\left.{\rm Trln}N^{-1}\right\vert^{div}=
\frac{iV_{4}}{(4\pi r^{2})^{2}\epsilon}\left[-\frac{4640}{189}\right].
\end{equation}
For $N=6$,
\begin{equation}
\left.{\rm Trln}N^{-1}\right\vert^{div}=
\frac{iV_{4}}{(4\pi r^{2})^{2}\epsilon}
\left[-\frac{1232816}{10395}\right].
\end{equation}

Similarly for Trln${\cal G}^{-1}$, we have for $N=2$,
\begin{equation}
\left.{\rm Trln}{\cal G}^{-1}\right\vert^{div}=
\frac{iV_{4}}{(4\pi r^{2})^{2}\epsilon}
\left[\frac{106}{15}-\frac{257}{10}(\Lambda r^{2})+
36(\Lambda r^{2})^{2}-\frac{189}{8}(\Lambda r^{2})^{3}\right].
\end{equation}
For $N=4$,
\begin{equation}
\left.{\rm Trln}{\cal G}^{-1}\right\vert^{div}=
\frac{iV_{4}}{(4\pi r^{2})^{2}\epsilon}
\left[-\frac{7574}{45}+\frac{67552}{315}(\Lambda r^{2})-
\frac{13504}{135}(\Lambda r^{2})^{2}+
\frac{1664}{81}(\Lambda r^{2})^{3}-
\frac{128}{81}(\Lambda r^{2})^{4}\right].
\end{equation}
For $N=6$,
\begin{eqnarray}
\left.{\rm Trln}{\cal G}^{-1}\right\vert^{div}=
\frac{iV_{4}}{(4\pi r^{2})^{2}\epsilon}
&&\left[\frac{5833069}{3360}-\frac{26312047}{24192}(\Lambda r^{2})+
\frac{6498535}{24192}(\Lambda r^{2})^{2}-
\frac{8375}{256}(\Lambda r^{2})^{3}\right.
\nonumber\\
&&\ \ \left.
+\frac{18125}{9216}(\Lambda r^{2})^{4}-
\frac{6875}{147456}(\Lambda r^{2})^{5}\right].
\end{eqnarray}

Next, for the operator $U_{1}$, one can evaluate the trace of some
general power of the operator. The results are the following.
For $N=2$,
\begin{eqnarray}
\left.{\rm Tr}U_{1}\right\vert^{div}&=&
\frac{iV_{4}}{(4\pi r^{2})^{2}\epsilon}
\left[\frac{148}{15}-\frac{26}{5}(\Lambda r^{2})\right],
\\
\left.{\rm Tr}U_{1}^{2}\right\vert^{div}&=&
\frac{iV_{4}}{(4\pi r^{2})^{2}\epsilon}
\left[24-40(\Lambda r^{2})+18(\Lambda r^{2})^{2}\right],
\\
\left.{\rm Tr}U_{1}^{3}\right\vert^{div}&=&
\frac{iV_{4}}{(4\pi r^{2})^{2}\epsilon}
\left[\frac{57}{5}-\frac{69}{2}(\Lambda r^{2})+
\frac{81}{2}(\Lambda r^{2})^{2}-\frac{81}{4}(\Lambda r^{2})^{3}
\right].
\end{eqnarray}
For $N=4$,
\begin{eqnarray}
\left.{\rm Tr}U_{1}\right\vert^{div}&=&
\frac{iV_{4}}{(4\pi r^{2})^{2}\epsilon}
\left[\frac{13432}{45}-\frac{95504}{2835}(\Lambda r^{2})\right],
\\
\left.{\rm Tr}U_{1}^{2}\right\vert^{div}&=&
\frac{iV_{4}}{(4\pi r^{2})^{2}\epsilon}
\left[\frac{6908}{5}-
\frac{3808}{9}(\Lambda r^{2})+
\frac{13312}{405}(\Lambda r^{2})^{2}\right],
\\
\left.{\rm Tr}U_{1}^{3}\right\vert^{div}&=&
\frac{iV_{4}}{(4\pi r^{2})^{2}\epsilon}
\left[\frac{9612}{5}-996(\Lambda r^{2})+
\frac{1600}{9}(\Lambda r^{2})^{2}-
\frac{896}{81}(\Lambda r^{2})^{3}
\right],
\\
\left.{\rm Tr}U_{1}^{4}\right\vert^{div}&=&
\frac{iV_{4}}{(4\pi r^{2})^{2}\epsilon}
\left[\frac{29412}{35}-
\frac{3136}{5}(\Lambda r^{2})+
\frac{2752}{15}(\Lambda r^{2})^{2}\right.
\nonumber\\
&&\ \ \left.-
\frac{2048}{81}(\Lambda r^{2})^{3}+
\frac{1024}{729}(\Lambda r^{2})^{4}
\right].
\end{eqnarray}
For $N=6$,
\begin{eqnarray}
\left.{\rm Tr}U_{1}\right\vert^{div}&=&
\frac{iV_{4}}{(4\pi r^{2})^{2}\epsilon}
\left[\frac{1670863}{945}-
\frac{67301}{756}(\Lambda r^{2})\right],
\\
\left.{\rm Tr}U_{1}^{2}\right\vert^{div}&=&
\frac{iV_{4}}{(4\pi r^{2})^{2}\epsilon}
\left[\frac{803450}{63}-
\frac{101855}{63}(\Lambda r^{2})+
\frac{77815}{1512}(\Lambda r^{2})^{2}\right],
\\
\left.{\rm Tr}U_{1}^{3}\right\vert^{div}&=&
\frac{iV_{4}}{(4\pi r^{2})^{2}\epsilon}
\left[\frac{193775}{6}-
\frac{53575}{8}(\Lambda r^{2})+
\frac{22475}{48}(\Lambda r^{2})^{2}-
\frac{2125}{192}(\Lambda r^{2})^{3}
\right],
\\
\left.{\rm Tr}U_{1}^{4}\right\vert^{div}&=&
\frac{iV_{4}}{(4\pi r^{2})^{2}\epsilon}
\left[\frac{2197000}{63}-
\frac{91375}{9}(\Lambda r^{2})+
1125(\Lambda r^{2})^{2}\right.
\nonumber\\
&&\ \ \left.-
\frac{8125}{144}(\Lambda r^{2})^{3}+
\frac{625}{576}(\Lambda r^{2})^{4}
\right],
\\
\left.{\rm Tr}U_{1}^{5}\right\vert^{div}&=&
\frac{iV_{4}}{(4\pi r^{2})^{2}\epsilon}
\left[\frac{41524375}{3024}-
\frac{996875}{192}(\Lambda r^{2})+
\frac{1611875}{2016}(\Lambda r^{2})^{2}-
\frac{18125}{288}(\Lambda r^{2})^{3}\right.
\nonumber\\
&&\ \ \left.
+\frac{15625}{6144}(\Lambda r^{2})^{4}-
\frac{3125}{73728}(\Lambda r^{2})^{5}
\right].
\end{eqnarray}

For traces involving the operator $U_{2}$, the calculation is
more complicated. Following the same procedure as above, we obtain,
for $N=2$,
\begin{eqnarray}
\left.{\rm Tr}U_{2}\right\vert^{div}&=&
\frac{iV_{4}}{(4\pi r^{2})^{2}\epsilon}
\left[10+
\frac{1}{2}(\Lambda r^{2})-
\frac{45}{2}(\Lambda r^{2})^{2}+
\frac{81}{4}(\Lambda r^{2})^{3}\right],
\\
\left.{\rm Tr}U_{1}U_{2}\right\vert^{div}&=&
\frac{iV_{4}}{(4\pi r^{2})^{2}\epsilon}
\left[\frac{151}{12}-
\frac{73}{2}(\Lambda r^{2})+
\frac{81}{2}(\Lambda r^{2})^{2}-
\frac{81}{4}(\Lambda r^{2})^{3}\right].
\end{eqnarray}
For $N=4$,
\begin{eqnarray}
\left.{\rm Tr}U_{2}\right\vert^{div}&=&
\frac{iV_{4}}{(4\pi r^{2})^{2}\epsilon}
\left[\frac{1724}{5}-
\frac{328}{3}(\Lambda r^{2})+
\frac{20432}{405}(\Lambda r^{2})^{2}\right.
\nonumber\\
&&\ \ \left.-
\frac{128}{9}(\Lambda r^{2})^{3}+
\frac{1024}{729}(\Lambda r^{2})^{4}\right],
\\
\left.{\rm Tr}U_{1}U_{2}\right\vert^{div}&=&
\frac{iV_{4}}{(4\pi r^{2})^{2}\epsilon}
\left[\frac{5211}{5}-
\frac{4976}{15}(\Lambda r^{2})-
\frac{176}{15}(\Lambda r^{2})^{2}\right.
\nonumber\\
&&\ \ \left.+
\frac{128}{9}(\Lambda r^{2})^{3}-
\frac{1024}{729}(\Lambda r^{2})^{4}\right],
\\
\left.{\rm Tr}U_{1}^{2}U_{2}\right\vert^{div}&=&
\frac{iV_{4}}{(4\pi r^{2})^{2}\epsilon}
\left[\frac{4419}{5}-
\frac{1936}{3}(\Lambda r^{2})+
\frac{5008}{27}(\Lambda r^{2})^{2}\right.
\nonumber\\
&&\ \ \left.-
\frac{2048}{81}(\Lambda r^{2})^{3}+
\frac{1024}{729}(\Lambda r^{2})^{4}\right],
\\
\left.{\rm Tr}U_{2}^{2}\right\vert^{div}&=&
\frac{iV_{4}}{(4\pi r^{2})^{2}\epsilon}
\left[\frac{4657}{5}-
\frac{9952}{15}(\Lambda r^{2})+
\frac{25312}{135}(\Lambda r^{2})^{2}\right.
\nonumber\\
&&\ \ \left.-
\frac{2048}{81}(\Lambda r^{2})^{3}+
\frac{1024}{729}(\Lambda r^{2})^{4}\right].
\end{eqnarray}
For $N=6$,
\begin{eqnarray}
\left.{\rm Tr}U_{2}\right\vert^{div}&=&
\frac{iV_{4}}{(4\pi r^{2})^{2}\epsilon}
\left[\frac{2356505}{2016}+
\frac{10560265}{32256}(\Lambda r^{2})-
\frac{12249815}{96768}(\Lambda r^{2})^{2}\right.
\nonumber\\
&&\ \ \left.
+\frac{717125}{36864}(\Lambda r^{2})^{3}-
\frac{26875}{18432}(\Lambda r^{2})^{4}+
\frac{3125}{73728}(\Lambda r^{2})^{5}\right],
\\
\left.{\rm Tr}U_{1}U_{2}\right\vert^{div}&=&
\frac{iV_{4}}{(4\pi r^{2})^{2}\epsilon}
\left[\frac{26262425}{2304}-
\frac{2887775}{1536}(\Lambda r^{2})+
\frac{3084025}{18432}(\Lambda r^{2})^{2}\right.
\nonumber\\
&&\ \ \left.
-\frac{230125}{12288}(\Lambda r^{2})^{3}+
\frac{26875}{18432}(\Lambda r^{2})^{4}-
\frac{3125}{73728}(\Lambda r^{2})^{5}\right],
\\
\left.{\rm Tr}U_{1}^{2}U_{2}\right\vert^{div}&=&
\frac{iV_{4}}{(4\pi r^{2})^{2}\epsilon}
\left[\frac{3026375}{144}-
\frac{39073375}{8064}(\Lambda r^{2})+
\frac{9993125}{32256}(\Lambda r^{2})^{2}\right.
\nonumber\\
&&\ \ \left.
+\frac{43625}{6144}(\Lambda r^{2})^{3}-
\frac{26875}{18432}(\Lambda r^{2})^{4}+
\frac{3125}{73728}(\Lambda r^{2})^{5}\right],
\\
\left.{\rm Tr}U_{1}^{3}U_{2}\right\vert^{div}&=&
\frac{iV_{4}}{(4\pi r^{2})^{2}\epsilon}
\left[\frac{5418875}{384}-
\frac{3549125}{672}(\Lambda r^{2})+
\frac{7434625}{9216}(\Lambda r^{2})^{2}\right.
\nonumber\\
&&\ \ \left.
-\frac{387875}{6144}(\Lambda r^{2})^{3}+
\frac{15625}{6144}(\Lambda r^{2})^{4}-
\frac{3125}{73728}(\Lambda r^{2})^{5}\right],
\\
\left.{\rm Tr}U_{2}^{2}\right\vert^{div}&=&
\frac{iV_{4}}{(4\pi r^{2})^{2}\epsilon}
\left[\frac{609125}{96}+
\frac{998125}{1536}(\Lambda r^{2})-
\frac{2394875}{4608}(\Lambda r^{2})^{2}\right.
\nonumber\\
&&\ \ \left.
+\frac{163625}{2304}(\Lambda r^{2})^{3}-
\frac{36875}{9216}(\Lambda r^{2})^{4}+
\frac{3125}{36864}(\Lambda r^{2})^{5}\right],
\\
\left.{\rm Tr}U_{1}U_{2}^{2}\right\vert^{div}&=&
\frac{iV_{4}}{(4\pi r^{2})^{2}\epsilon}
\left[\frac{117011375}{8064}-
\frac{38508875}{7168}(\Lambda r^{2})+
\frac{26252375}{32256}(\Lambda r^{2})^{2}\right.
\nonumber\\
&&\ \ \left.
-\frac{583625}{9216}(\Lambda r^{2})^{3}+
\frac{15625}{6144}(\Lambda r^{2})^{4}-
\frac{3125}{73728}(\Lambda r^{2})^{5}\right].
\end{eqnarray}

Putting all these together, we can obtain the divergent parts
of the VD effective actions for various $M^{4}\times S^{N}$
Kaluza-Klein spaces. For $M^{4}\times S^{2}$, we have
\begin{eqnarray}
iW^{div}_{unique}&=&\left.
{\rm Trln}N^{-1}-\frac{1}{2}{\rm Trln}{\cal G}^{-1}+
\frac{1}{2}{\rm Tr}U_{1}+\frac{1}{4}{\rm Tr}U^{2}_{1}-
\frac{1}{2}{\rm Tr}U_{2}+\frac{1}{6}{\rm Tr}U^{3}_{1}-
\frac{1}{2}{\rm Tr}U_{1}U_{2}\right\vert^{div}
\nonumber\\
&=&
\frac{iV_{4}}{(4\pi r^{2})^{2}\epsilon}
\left[-\frac{2249}{840}+
\frac{25}{2}(\Lambda r^{2})-
\frac{63}{4}(\Lambda r^{2})^{2}+
\frac{135}{16}(\Lambda r^{2})^{3}
\right].
\label{w2}
\end{eqnarray}
This result is consistent with that in ref.\cite{CK1} where the 
divergent part of the VD effective action was calculated in a 
general six-dimensional background spacetime. 
For $M^{4}\times S^{4}$,
\begin{eqnarray}
iW^{div}_{unique}&=&
{\rm Trln}N^{-1}-\frac{1}{2}{\rm Trln}{\cal G}^{-1}+
\frac{1}{2}{\rm Tr}U_{1}+\frac{1}{4}{\rm Tr}U^{2}_{1}-
\frac{1}{2}{\rm Tr}U_{2}+\frac{1}{6}{\rm Tr}U^{3}_{1}-
\frac{1}{2}{\rm Tr}U_{1}U_{2}
\nonumber\\
&&\ \ \left.+\frac{1}{8}{\rm Tr}U^{4}_{1}-
\frac{1}{2}{\rm Tr}U^{2}_{1}U_{2}+
\frac{1}{4}{\rm Tr}U^{2}_{2}\right\vert^{div}
\nonumber\\
&=&
\frac{iV_{4}}{(4\pi r^{2})^{2}\epsilon}
\left[\frac{41657}{540}-
\frac{7850}{81}(\Lambda r^{2})+
\frac{6152}{135}(\Lambda r^{2})^{2}\right.
\nonumber\\
&&\ \ \left.-
\frac{2176}{243}(\Lambda r^{2})^{3}+
\frac{448}{729}(\Lambda r^{2})^{4}
\right].
\label{w4}
\end{eqnarray}
Finally for $M^{4}\times S^{6}$ we obtain,
\begin{eqnarray}
iW^{div}_{unique}&=&
{\rm Trln}N^{-1}-\frac{1}{2}{\rm Trln}{\cal G}^{-1}+
\frac{1}{2}{\rm Tr}U_{1}+\frac{1}{4}{\rm Tr}U^{2}_{1}-
\frac{1}{2}{\rm Tr}U_{2}+\frac{1}{6}{\rm Tr}U^{3}_{1}-
\frac{1}{2}{\rm Tr}U_{1}U_{2}
\nonumber\\
&&\ \ \left.+\frac{1}{8}{\rm Tr}U^{4}_{1}
-\frac{1}{2}{\rm Tr}U^{2}_{1}U_{2}+
\frac{1}{4}{\rm Tr}U^{2}_{2}+
\frac{1}{10}{\rm Tr}U^{5}_{1}-
\frac{1}{2}{\rm Tr}U^{3}_{1}U_{2}+
\frac{1}{2}{\rm Tr}U_{1}U^{2}_{2}\right\vert^{div}
\nonumber\\
&=&\frac{iV_{4}}{(4\pi r^{2})^{2}\epsilon}
\left[-\frac{476483023}{591360}+
\frac{10896475}{21504}(\Lambda r^{2})-
\frac{32112235}{258048}(\Lambda r^{2})^{2}\right.
\nonumber\\
&&\ \ \ \ \ \left.
+\frac{549625}{36864}(\Lambda r^{2})^{3}-
\frac{10625}{12288}(\Lambda r^{2})^{4}+
\frac{625}{32768}(\Lambda r^{2})^{5}\right].
\label{w6}
\end{eqnarray}

\section{Applications}

In this section we make two applications of the above effective actions. 
We give trace anomalies and we search for self-consistent
configurations of the internal Kaluza-Klein spheres. 

Some explanation of a trace anomaly is in order for gravity. 
Einstein gravity is
not Weyl invariant and the trace part we calculate is
NOT a combination of the anomalous and the ``normal" contributions. What
we calculate below is the ``anomalous" part of the trace, as discussed
in previous works. In ``Trace anomaly for gravitons" \cite{RC} Critchley
argues that the ``anomalous" part of the trace is given by $a_2$, while the first terms of
his (4) gives the ``normal" contribution, irrespective to the value of $\zeta$.
For Einstein gravity, this is equivalent
to his (48), which is the same as our calculation. It is in this sense that we
calculate the ``anomalous" part of the trace anomaly. This is not the same
as the usual trace anomaly in which the
original classical action is conformal, while the quantum action is not.
For nonconformal theories, Duff \cite{MJD} in his (29)
gave a definition for the anomaly, which we suspect is the same as what we have
calculated.

This problem is also discussed in ``Quantum fields in curved space"
\cite{BiD}. On page 179, the authors mention
that for fields which are not conformally invariant, there will be extra
non-anomalous (normal) terms, which are really the same as the first terms
of (4) in Critchley's paper.

The paper ``Non-conformal renormalized stress tensors in Robertson-Walker space-times", 
\cite{BuD} by Bunch and Davies,
gave a rather detailed calculation for the stress tensor for a massless,
but minimally coupled scalar (therefore a non-conformally invariant field) in a conformal
Robertson-Walker spacetime. Their (3.16) gives the total trace of the
stress tensor of the quantum theory and (3.18) gives the ``normal" part
of the trace. Subtracting (3.18) from (3.16) (using Duff's definition) gives
(3.19), the ``anomalous" trace part which should be equivalent to
what we have calculated because (3.19) is proportional to $a_2$. 
                
In the spirit of the above we can easily determine the gauge-independent VD trace
anomaly for gravitons from the divergent part of the effective action of
the corresponding Kaluza-Klein
space. The appropriate expression is:
\begin{eqnarray}
\langle T^{\mu}_{\mu}\rangle_{ren}
&=&\frac{\epsilon W^{div}_{unique}}{V_{4}V_{S^{N}}},
\nonumber\\
{\rm where \hskip .5 in}\  V_{S^{N}}&=& 
{2\pi^{(N+1)/2}r^{N}\over\Gamma\left(\frac{N+1}{2}\right) },
\label{VSN}
\end{eqnarray}
is the volume of the sphere $S^N$.
Thus, for $M^{4}\times S^{2}$, 
\begin{equation}
\langle T^{\mu}_{\mu}\rangle_{ren}=
\frac{1}{(4\pi)^{3}}\left[
\frac{135}{16}\Lambda^{3}-
\frac{63}{4}\left(\frac{\Lambda^{2}}{r^{2}}\right)+
\frac{25}{2}\left(\frac{\Lambda}{r^{4}}\right)-
\frac{2249}{840}\left(\frac{1}{r^{6}}\right)\right],
\end{equation}
for $M^{4}\times S^{4}$,
\begin{equation}
\langle T^{\mu}_{\mu}\rangle_{ren}=
\frac{1}{(4\pi)^{4}}\left[
\frac{896}{243}\Lambda^{4}-
\frac{4352}{81}\left(\frac{\Lambda^{3}}{r^{2}}\right)+
\frac{12304}{45}\left(\frac{\Lambda^{2}}{r^{4}}\right)-
\frac{15700}{27}\left(\frac{\Lambda}{r^{6}}\right)+
\frac{41657}{90}\left(\frac{1}{r^{8}}\right)\right],
\end{equation}
and for $M^{4}\times S^{6}$,
\begin{eqnarray}
\langle T^{\mu}_{\mu}\rangle_{ren}&=&
\frac{1}{(4\pi)^{5}}\left[
\frac{9375}{8192}\Lambda^{5}-
\frac{53125}{1024}\left(\frac{\Lambda^{4}}{r^{2}}\right)+
\frac{336875}{384}\left(\frac{\Lambda^{3}}{r^{4}}\right)-
\frac{160561175}{21504}\left(\frac{\Lambda^{2}}{r^{6}}\right)\right.
\nonumber\\
&&\ \ \left.
+\frac{54482375}{1792}\left(\frac{\Lambda}{r^{8}}\right)-
\frac{476483023}{9856}\left(\frac{1}{r^{10}}\right)\right].
\end{eqnarray}
The pure $\Lambda$ terms in all three cases can be compared with 
\cite{CK3} and the N = 2 case agrees with \cite{CK1}.

Our second use of the VD effective actions computed above is to produce stable 
configurations of the internal spheres. Because quantum fluctuations of the gravity field itself (about the Kaluza-Klein 
background) are generating the corrections to the effective potential, 
any stable configuration should
be called a self-consistent dimensionally reduced configuration. For a stable
configuration to
be of interest it should have a  positive renormalized Newton's constant $G_0$. 
Most efforts to find such configurations have 
used the naive effective action (see \cite{RDS,CM,M}) and have failed. 
Stable configurations 
were found but
only with  negative gravity constants. Those efforts that attempted
to correctly use  the VD
effective potential have not gotten past the simplest cases: $M^{4}\times S^{1}$ or 
$M^{4}\times S^{2},$ and $M^{4}\times S^{1}\times S^{1}$.  
For these cases no acceptable configurations were 
found either \cite{BO,HKLT,BLO,SDOb,BKLM,BKO,SDO} .

Because of our lack of knowledge of the finite part of the 
effective potential we will assume that the divergent part ($\propto 1/\epsilon$) 
dominates at one-loop. We 
can then easily seek stable configurations for the internal geometry, i.e., configurations 
where the classical gravity forces balance the quantum gravity 
(Casimir) pressures.
We follow \cite{CW} and seek configurations satisfying
$$V(r)={\partial V \over \partial r} =0, \quad {\partial^2 V \over 
\partial r^2} > 0,$$
where the potential $V(r)$ is the negative of the 1-loop corrected 
effective action:
\begin{equation} 
-V(r)\equiv {1\over 16\pi G}\left({N(N-1)\over r^2}-2\Lambda\right)+
{W_{unique}^{div}\over V_4}. 
\label{efV1}
\end{equation}
The bare value of Newton's 4-dimensional gravity parameter G depends 
inversely on the sphere's volume $V_{S^N}$ 
[see (\ref{VSN})], i.e., $G\times V_{S^N}$ is the initial gravity 
\underbar{constant} in 4+N dimensions. 
Values for $W_{unique}^{div}$ can be found in (\ref{w2}), (\ref{w4}), 
and  (\ref{w6}).

There are only two static configurations for even dimensions $\le 6$  
(see columns 2 and 3 of Table I), one stable and one not. 

\begin{table}
\caption{The only 2 static configurations for
$M^4\times S^N$ (N = 2, 4, or 6) at 1-loop. Columns 2 and 3 are required by
stability ($\partial^2 V/\partial r^2>0$), columns 4 and 5 relate the bare parameters G and $\Lambda$ to the 
renormalized parameters G$_0$ and $\Lambda_0$, and columns 6-8 
are resulting parameter values computed using columns 2-5. }
\begin{tabular}{c|cc|cc|ccc}
\hline
N & $\Lambda r^2$ & $ G/ \epsilon r^2$  &
G/G$_0$ & $\Lambda/\Lambda_0$ & $r^2\Lambda_0$ & $G_0\Lambda_0/
\epsilon$ & $g^2/G_0\Lambda_0$\\
\hline
2 & 2.3609 & 0.17076  & -1.3857 & 4.9834 & 0.47375 & $-$0.058377 & 159.15\\
6 & 5.6876 & $-$22.931  & 221.15 & 2.9948 & 1.8991 & $-$0.19692 & 92.636\\
\hline
\end{tabular}
\end{table}
 
Columns 4 and 5 relate the bare parameters to their renormalized 
values. 
Because Einstein gravity is not renormalizable we are rather  
unconstrained in our use of the effective actions (potentials).
We have done the 1-loop renormalization by rewriting (\ref{efV1}) as
\begin{eqnarray} 
-V(r)&=& {1\over 16\pi G}\left({N(N-1)\over r^2}-2\Lambda\right)
\nonumber\\
&+&{1 \over (4\pi)^2\epsilon r^4}\left\{ 
c_0+c_1(\Lambda r^2)+c_2(\Lambda r^2)^2+\cdots+c_{N/2+2}(\Lambda r^2)^{N/2+2}\right\}, 
\label{efV2}
\end{eqnarray}
and collecting the $r^{N}$ and $r^{N-2}$ terms to get two equations:
\begin{equation} 
{1\over 16\pi G}\left(-2\Lambda\right) +{1 \over (4\pi)^2\epsilon}
\left\{c_{N/2+2} \Lambda^{N/2+2}r^N \right\}=
{1\over 16\pi G_0}\left(-2\Lambda_0\right),
\label{c2}
\end{equation}
\begin{equation} 
{1\over 16\pi G}\left({N(N-1)\over r^2}\right) 
+{1 \over (4\pi)^2\epsilon}\left\{c_{N/2+1} \Lambda^{N/2+1}r^{N-2} \right\}=
{1\over 16\pi G_0}\left({N(N-1)\over r^2}\right).
\label{c1}
\end{equation}
The constants 
$c_{N/2+1}$ and $c_{N/2+2}$ can be read from (\ref{w2}),(\ref{w4}), and (\ref{w6}).
Solving these two equations gives:

\begin{equation} 
{G\over G_0}=1+c_{N/2+1}\left(\Lambda r^2\right)^{N/2+1} 
(G/\pi\epsilon r^2)/N(N-1),
\label{G}
\end{equation}
\begin{equation} 
{\Lambda\over\Lambda_0}={
1+c_{N/2+1}\left(\Lambda r^2\right)^{N/2+1} 
(G/\pi\epsilon r^2)/N(N-1)
\over
1-c_{N/2+2}\left(\Lambda r^2\right)^{N/2+1} 
(G/\pi\epsilon r^2)/2.
}
\label{Lambda}
\end{equation}

Both stationary configurations in the Table have negative values for the 
external dimensional regularization parameter 
$\epsilon=4-d$ and positive values for the renormalized Newton's and cosmological 
constants G$_0$ and $\Lambda_0$. 
The $N=2$ configuration is stable but requires a negative bare 
gravity constant G. The $N=6$ configuration has
a positive unrenormalized gravity constant G, but is unstable.

Two input parameters, e.g., $G_0$ and $\Lambda_0$, are required to evaluate all 
other parameters (see columns 6-8) including the renormalized $O(N+1)$ 
coupling constant $g^2=(N+1)8\pi G_0/r^2$, see \cite{CW}. As expected, this theory 
cannot apply to the current phase of the universe where $\Lambda_0\le 10^{-52} m^{-2}$
and $G_0= 2.6\times 10^{-70} m^2$, but perhaps to one of the 
pre-inflation phases where $G_0\Lambda_0\sim 10^{-2}$.

\section{Conclusions}

We have demonstrated how to evaluate the divergent part of the VD
effective action for Einstein gravity on even-dimensional Kaluza-Klein spacetimes 
of the form $M^{4}\times S^{N}$.
First the effective action is expanded as a series of functional
traces of various operators including $N^{-1}$, ${\cal G}^{-1}$, 
$U_{1}$, and $U_{2}$ and some of their products. Then the eigenvalues of these operators
are obtained by decomposing the corresponding eigenfunctions, 
vectors or symmetric tensors, into their irreducible parts. 
Using these eigenvalues, the divergent parts of the traces can
be obtained, thus giving the divergent parts 
of the VD effective action. 
The formulae used to extract these divergent
parts are tabulated in the Appendix for $N$=2, 4, and 6.

Although the above procedure becomes more and more tedious as
one goes to higher dimensions, there is no conceptual 
difficulty in doing so. One can therefore extend this method
to even dimensions with $N\ge 8$, as well as to other more general coset 
spaces (provided eigenvalues
for the corresponding Laplacians are known).

From the divergent parts of the VD effective action, we 
have obtained the gauge-independent trace anomaly
for gravitons on $M^{4}\times S^{N}$. The trace anomaly
for gravitons derived from the usual effective action depends
on the choice of gauge condition in the off-shell case; however, 
the VD formalism provides an alternative definition
of a unique trace anomaly, even when 
off-shell \cite{CK3}. Our final application was an attempt to find 
self-consistent dimensionally reduced Kaluza-Klein spaces.
Only one was found (N=2) and it required that we start with
a negative bare Newton's constant. It also possessed much too
small of a gauge coupling constant to represent the current 
stage of the universe.

\begin{acknowledgments}
H. T. Cho is supported by the National Science Council
of the Republic of China under contract number 
NSC 87-2112-M-032-001. 
\end{acknowledgments}

\appendix
\section*{}

In this appendix we evaluate the divergent part of the traces
of various operators. We use the dimensional regularization
scheme in which the dimension of the external spacetime is taken
to be
\begin{equation}
d\rightarrow 4-\epsilon.
\end{equation}
Then we extract the part which is proportional to 
$1/\epsilon$ as $\epsilon\rightarrow 0$. 

The traces that we shall consider are
\begin{eqnarray}
&&\sum_{k}{\rm ln}k^{2},
\nonumber\\
&&\sum_{k}\left(\frac{1}{k^{2}}\right)^{p},
\nonumber\\
&&\sum_{k}\sum^{\infty}_{l=0}{D^{(s,v,t)}_{l}}(N)
{\rm ln}(k^{2}-\Lambda_{l}(N))\equiv
\frac{iV_{4}}{(4\pi r^{2})^{2}\epsilon}F^{(s,v,t)}(N),
\nonumber\\
&&\sum_{k}\sum^{\infty}_{l=0}D^{(s,v,t)}(N)(k^{2})^{p}
\left(\frac{1}{k^{2}-\Lambda_{l}(N)}\right)^{q}\equiv
\frac{iV_{4}}{(4\pi r^{2})^{2}\epsilon}G^{(s,v,t)}(p,q,N),
\end{eqnarray}
for $p\geq 0$, $q\geq 1$.

First, using the proper time method, the divergent part of 
$\sum_{k}{\rm ln}k^{2}$ can be written as
\begin{eqnarray}
\left.\sum_{k}{\rm ln}k^{2}\right\vert^{div}
&=&\left.-\int^{\infty}_{0}\frac{d\tau}{\tau}
\sum_{k}e^{-\tau k^{2}}e^{-\tau m^{2}}
\right\vert^{div}_{m\rightarrow 0}
\nonumber\\
&=&\left.-\frac{iV_{d}}{(4\pi)^{d/2}}\int^{\infty}_{0}
d\tau\ \tau^{-d/2-1}e^{-\tau m^{2}}
\right\vert^{div}_{m\rightarrow 0}
\nonumber\\
&=&\left.-\frac{iV_{d}}{(4\pi)^{2}}\frac{1}{\epsilon}(m^{4})
\right\vert_{m\rightarrow 0}
\nonumber\\
&=&0.
\end{eqnarray}

Similarly,
\begin{equation}
\left.\sum_{k}\left(\frac{1}{k^{2}}\right)^{p}\right\vert^{div}
=\frac{iV_{4}}{(4\pi)^{2}}\frac{1}{\epsilon}(2)\delta_{p2},
\end{equation}
that is, the trace has a divergent part only when $p=2$.

Next, we consider $F^{(s,v,t)}(N)$. For $N=2$, $D^{(t)}_{l}=0$ for
$l\geq 1$ and $D^{(t)}_{l=0}=-3$. While
\begin{equation}
D^{(v)}_{l}=D^{(s)}_{l}=2l+1,
\end{equation}
and
\begin{equation}
\Lambda_{l}(2)=-\frac{l(l+1)}{r^{2}},
\end{equation}
for $l\geq 0$. Therefore,
\begin{eqnarray}
\left.\sum_{k}\sum^{\infty}_{l=0}D^{(t)}_{l}(2){\rm ln}
(k^{2}-\Lambda_{l}(2))\right\vert^{div} 
&=&\left.\sum_{k}(-3){\rm ln}k^{2}\right\vert^{div}
\nonumber\\
&=&0,
\end{eqnarray}
and 
\begin{eqnarray}
&&\left.\sum_{k}\sum^{\infty}_{l=0}D^{(v)}_{l}(2)
{\rm ln}(k^{2}-\Lambda_{l}(2))\right\vert^{div}
\nonumber\\
&=&\left.\sum_{k}\sum^{\infty}_{l=0}(2l+1){\rm ln}
\left(k^{2}+\frac{l(l+1)}{r^{2}}\right)\right\vert^{div}
\nonumber\\
&=&\left.-\int^{\infty}_{0}\frac{d\tau}{\tau}
\left(\sum_{k}e^{-\tau k^{2}}\right)
\left(\sum^{\infty}_{l=0}(2l+1)e^{-\tau l(l+1)/r^{2}}\right)
e^{-\tau m^{2}}\right\vert^{div}_{m\rightarrow 0}
\nonumber\\
&=&\left.-\int^{\infty}_{0}\frac{d\tau}{\tau}
\frac{iV_{d}}{(4\pi)^{d/2}}
\frac{e^{-\tau m^{2}}}{\tau^{d/2}}
\left(\frac{r^{2}}{\tau}+\frac{1}{3}+
\frac{\tau}{15r^{2}}+\frac{4\tau^{2}}{315r^{4}}+\cdots\right)
\right\vert^{div}_{m\rightarrow 0}
\nonumber\\
&=&\frac{iV_{4}}{(4\pi r^{2})^{2}}\frac{1}{\epsilon}
\left(-\frac{8}{315}\right).
\end{eqnarray}
Note that we have used an asymptotic expansion for the summation
over $l$ above for small values of $\tau$ \cite{BKM}.
Now we have
\begin{eqnarray}
F^{(t)}(2)&=&0,
\\
F^{(v)}(2)&=&F^{(s)}(2)=-\frac{8}{315}.
\end{eqnarray}

For $N=4$,
\begin{eqnarray}
D^{(t)}_{l}&=&\frac{5}{6}(2l+3)(l+4)(l-1),
\\
D^{(v)}_{l}&=&\frac{1}{2}(2l+3)(l+3)l,
\\
D^{(s)}_{l}&=&\frac{1}{6}(2l+3)(l+2)(l+1),
\\
\Lambda_{l}&=&-\frac{l(l+3)}{r^{2}},
\end{eqnarray}
for $l\geq 0$, and the asymptotic expansions,
\begin{eqnarray}
&&\sum^{\infty}_{l=0}\frac{5}{6}(2l+3)(l+4)(l-1)
e^{-\tau l(l+3)/r^{2}}
\nonumber\\
&=&\frac{5r^{4}}{6\tau^{2}}-\frac{10r^{2}}{3\tau}-
\frac{91}{18}-\frac{508\tau}{189r^{2}}-
\frac{127\tau^{2}}{378r^{4}}+\frac{806\tau^{3}}{2079r^{6}}+
\frac{21311\tau^{4}}{81081r^{8}}+
\frac{3416\tau^{5}}{57915r^{10}}+\cdots,
\\
&&\sum^{\infty}_{l=0}\frac{1}{2}(2l+3)(l+3)l
e^{-\tau l(l+3)/r^{2}}
\nonumber\\
&=&\frac{r^{4}}{2\tau^{2}}-\frac{11}{30}-
\frac{46\tau}{315r^{2}}+\frac{19\tau^{2}}{210r^{4}}+
\frac{388\tau^{3}}{3465r^{6}}+
\frac{6179\tau^{4}}{135135r^{8}}-
\frac{268\tau^{5}}{225225r^{10}}+\cdots,
\\
&&\sum^{\infty}_{l=0}\frac{1}{6}(2l+3)(l+2)(l+1)
e^{-\tau l(l+3)/r^{2}}
\nonumber\\
&=&\frac{r^{4}}{6\tau^{2}}+\frac{r^{2}}{3\tau}+
\frac{29}{90}+\frac{37\tau}{189r^{2}}+
\frac{149\tau^{2}}{1890r^{4}}+
\frac{179\tau^{3}}{10395r^{6}}-
\frac{1387\tau^{4}}{405405r^{8}}-
\frac{13162\tau^{5}}{2027025r^{10}}+\cdots.
\end{eqnarray}
We obtain
\begin{eqnarray}
F^{(t)}(4)&=&\frac{127}{189},
\\
F^{(v)}(4)&=&-\frac{19}{105},
\\
F^{(s)}(4)&=&-\frac{149}{945}.
\end{eqnarray}

For $N=6$,
\begin{eqnarray}
D^{(t)}_{l}&=&\frac{7}{60}(2l+5)(l+6)(l+3)(l+2)(l-1),
\\
D^{(v)}_{l}&=&\frac{1}{24}(2l+5)(l+5)(l+3)(l+2)l,
\\
D^{(s)}_{l}&=&\frac{1}{120}(2l+5)(l+4)(l+3)(l+2)(l+1),
\\
\Lambda_{l}&=&-\frac{l(l+5)}{r^{2}},
\end{eqnarray}
for $l\geq 0$, and the asymptotic expansions,
\begin{eqnarray}
&&\sum^{\infty}_{l=0}
\frac{7}{60}(2l+5)(l+6)(l+3)(l+2)(l-1)
e^{-\tau l(l+5)/r^{2}}
\nonumber\\
&=&\frac{7r^{6}}{30\tau^{3}}-\frac{21r^{2}}{5\tau}-
\frac{262}{27}-\frac{4133\tau}{450r^{2}}-
\frac{248\tau^{2}}{99r^{4}}+
\frac{958729\tau^{3}}{289575r^{6}}+
\frac{14624\tau^{4}}{3861r^{8}}
\nonumber\\
&&\ \ 
+\frac{11267779\tau^{5}}{16409250r^{10}}-
\frac{357108736\tau^{6}}{168358905r^{12}}+\cdots,
\\
&&\sum^{\infty}_{l=0}\frac{1}{24}(2l+5)(l+5)(l+3)(l+2)l
e^{-\tau l(l+5)/r^{2}}
\nonumber\\
&=&\frac{r^{6}}{12\tau^{3}}+
\frac{r^{4}}{4\tau^{2}}-
\frac{1823}{3780}-
\frac{487\tau}{1260r^{2}}+
\frac{3671\tau^{2}}{13860r^{4}}+
\frac{264611\tau^{3}}{405405r^{6}}+
\frac{1603\tau^{4}}{4290r^{8}}
\nonumber\\
&&\ \ -\frac{9333977\tau^{5}}{45945900r^{10}}-
\frac{13067106599\tau^{6}}{23570246700r^{12}}+\cdots,
\\
&&\sum^{\infty}_{l=0}\frac{1}{120}(2l+5)(l+4)(l+3)(l+2)(l+1)
e^{-\tau l(l+5)/r^{2}}
\nonumber\\
&=&\frac{r^{6}}{60\tau^{3}}+
\frac{r^{4}}{12\tau^{2}}+
\frac{r^{2}}{5\tau}+
\frac{1139}{3780}+
\frac{833\tau}{2700r^{2}}+
\frac{137\tau^{2}}{660r^{4}}+
\frac{121442\tau^{3}}{2027025r^{6}}-
\frac{45251\tau^{4}}{810810r^{8}}
\nonumber\\
&&\ \ -\frac{23068481\tau^{5}}{229729500r^{10}}-
\frac{1974977293\tau^{6}}{23570246700r^{12}}+\cdots.
\end{eqnarray}
We obtain
\begin{eqnarray}
F^{(t)}(6)&=&\frac{496}{99},
\\
F^{(v)}(6)&=&-\frac{3671}{6930},
\\
F^{(s)}(6)&=&-\frac{137}{330}.
\end{eqnarray}

For $G^{(s,v,t)}(p,q,N)$, we again use the asymptotic expansions 
for various dimensions and the same procedure to extract the 
divergent parts. The results are listed out in the following
tables.

\vspace{40pt}
\begin{tabular}{c|cccccc}
\hline\hline
$G^{(t)}(p,q,2)$ & $q=1$ & 2 & 3 & 4 & 5 & 6 \\
\hline
$p=0$ & 0 & $-6r^{4}$ & 0 & 0 & 0 & 0 \\
1 & 0 & 0 & $-6r^{4}$ & 0 & 0 & 0 \\
2 & 0 & 0 & 0 & $-6r^{4}$ & 0 & 0 \\
3 & 0 & 0 & 0 & 0 & $-6r^{4}$ & 0 \\ 
\hline\hline
\end{tabular}

\vspace{40pt}
\begin{tabular}{c|cccccc}
\hline\hline
$G^{(v)}(p,q,2)$ & $q=1$ & 2 & 3 & 4 & 5 & 6 \\
\hline
$p=0$ & $\frac{2r^{2}}{15}$ & $\frac{2r^{4}}{3}$ & $r^{6}$ 
& 0 & 0 & 0 \\
1 & $\frac{16}{315}$ & $\frac{4r^{2}}{15}$ & $\frac{2r^{4}}{3}$ 
& $\frac{2r^{6}}{3}$ & 0 & 0 \\
2 & $\frac{4}{105r^{2}}$ & $\frac{16}{105}$ & $\frac{2r^{2}}{5}$ 
& $\frac{2r^{4}}{3}$ & $\frac{r^{6}}{2}$ & 0 \\
3 & $\frac{64}{1155r^{4}}$ & $\frac{16}{105r^{2}}$  
& $\frac{32}{105}$ & $\frac{8r^{2}}{15}$ 
& $\frac{2r^{4}}{3}$ 
& $\frac{2r^{6}}{5}$ \\ 
\hline\hline
\end{tabular}

\vspace{40pt}
\begin{tabular}{c|cccccc}
\hline\hline
$G^{(s)}(p,q,2)$ & $q=1$ & 2 & 3 & 4 & 5 & 6 \\
\hline
$p=0$ & $\frac{2r^{2}}{15}$ & $\frac{2r^{4}}{3}$ & $r^{6}$ 
& 0 & 0 & 0 \\
1 & $\frac{16}{315}$ & $\frac{4r^{2}}{15}$ & $\frac{2r^{4}}{3}$ 
& $\frac{2r^{6}}{3}$ & 0 & 0 \\
2 & $\frac{4}{105r^{2}}$ & $\frac{16}{105}$ & $\frac{2r^{2}}{5}$ 
& $\frac{2r^{4}}{3}$ & $\frac{r^{6}}{2}$ & 0 \\
3 & $\frac{64}{1155r^{4}}$ & $\frac{16}{105r^{2}}$  
& $\frac{32}{105}$ & $\frac{8r^{2}}{15}$ 
& $\frac{2r^{4}}{3}$ 
& $\frac{2r^{6}}{5}$ \\ 
\hline\hline
\end{tabular}

\vspace{40pt}
\begin{tabular}{c|cccccccc}
\hline\hline
$G^{(t)}(p,q,4)$ & $q=1$ & 2 & 3 & 4 & 5 & 6 & 7 & 8 \\
\hline
$p=0$ & $-\frac{1016r^{2}}{189}$ & $-\frac{91r^{4}}{9}$ &
$-\frac{10r^{6}}{3}$ & $\frac{5r^{8}}{18}$ & 0 & 0 & 0 & 0 \\
1 & $-\frac{254}{189}$ & $-\frac{2032r^{2}}{189}$ &
$-\frac{91r^{4}}{9}$ & $-\frac{20r^{6}}{9}$ &
$\frac{5r^{8}}{36}$ & 0 & 0 & 0 \\
2 & $\frac{3224}{693r^{2}}$ & $-\frac{254}{63}$ &
$-\frac{1016r^{2}}{63}$ & $-\frac{91r^{4}}{9}$ &
$-\frac{5r^{6}}{3}$ & $\frac{r^{8}}{12}$ & 0 & 0 \\
3 & $\frac{340976}{27027r^{4}}$ & $\frac{12896}{693r^{2}}$ &
$-\frac{508}{63}$ & $-\frac{4064r^{2}}{189}$ &
$-\frac{91r^{4}}{9}$ & $-\frac{4r^{6}}{3}$ &
$\frac{r^{8}}{18}$ & 0 \\
4 & $\frac{54656}{3861r^{6}}$ & $\frac{1704880}{27027r^{4}}$ &
$\frac{32240}{693r^{2}}$ & $-\frac{2540}{189}$ &
$-\frac{5080r^{2}}{189}$ & $-\frac{91r^{4}}{9}$ &
$-\frac{10r^{6}}{9}$ & $\frac{5r^{8}}{126}$ \\
\hline\hline
\end{tabular}

\vspace{40pt}
\begin{tabular}{c|cccccccc}
\hline\hline
$G^{(v)}(p,q,4)$ & $q=1$ & 2 & 3 & 4 & 5 & 6 & 7 & 8 \\
\hline
$p=0$ & $-\frac{92r^{2}}{315}$ & $-\frac{11r^{4}}{15}$ &
0 & $\frac{r^{8}}{6}$ & 0 & 0 & 0 & 0 \\
1 & $\frac{38}{105}$ & $-\frac{184r^{2}}{315}$ &
$-\frac{11r^{4}}{15}$ & 0 &
$\frac{r^{8}}{12}$ & 0 & 0 & 0 \\
2 & $\frac{1552}{1155r^{2}}$ & $\frac{38}{35}$ &
$-\frac{92r^{2}}{105}$ & $-\frac{11r^{4}}{15}$ &
0 & $\frac{r^{8}}{20}$ & 0 & 0 \\
3 & $\frac{98864}{45045r^{4}}$ & $\frac{6208}{1155r^{2}}$ &
$-\frac{76}{35}$ & $-\frac{368r^{2}}{315}$ &
$-\frac{11r^{4}}{15}$ & 0 &
$\frac{r^{8}}{30}$ & 0 \\
4 & $-\frac{4288}{15015r^{6}}$ & $\frac{98864}{9009r^{4}}$ &
$\frac{3104}{231r^{2}}$ & $\frac{76}{21}$ &
$-\frac{92r^{2}}{63}$ & $-\frac{11r^{4}}{15}$ &
0 & $\frac{r^{8}}{42}$ \\
\hline\hline
\end{tabular}

\vspace{40pt}
\begin{tabular}{c|cccccccc}
\hline\hline
$G^{(s)}(p,q,4)$ & $q=1$ & 2 & 3 & 4 & 5 & 6 & 7 & 8 \\
\hline
$p=0$ & $\frac{74r^{2}}{189}$ & $\frac{29r^{4}}{45}$ &
$\frac{r^{6}}{3}$ & $\frac{r^{8}}{18}$ & 0 & 0 & 0 & 0 \\
1 & $\frac{298}{945}$ & $\frac{148r^{2}}{189}$ &
$\frac{29r^{4}}{45}$ & $\frac{2r^{6}}{9}$ &
$\frac{r^{8}}{36}$ & 0 & 0 & 0 \\
2 & $\frac{716}{3465r^{2}}$ & $\frac{298}{315}$ &
$\frac{74r^{2}}{63}$ & $\frac{29r^{4}}{45}$ &
$\frac{r^{6}}{6}$ & $\frac{r^{8}}{60}$ & 0 & 0 \\
3 & $-\frac{22192}{135135r^{4}}$ & $\frac{2864}{3465r^{2}}$ &
$\frac{596}{315}$ & $\frac{296r^{2}}{189}$ &
$\frac{29r^{4}}{45}$ & $\frac{2r^{6}}{15}$ &
$\frac{r^{8}}{90}$ & 0 \\
4 & $-\frac{210592}{135135r^{6}}$ & $-\frac{22192}{27027r^{4}}$ &
$\frac{1432}{693r^{2}}$ & $\frac{596}{189}$ &
$\frac{370r^{2}}{189}$ & $\frac{29r^{4}}{45}$ &
$\frac{r^{6}}{9}$ & $\frac{r^{8}}{126}$ \\
\hline\hline
\end{tabular}

\vspace{40pt}
\begin{tabular}{c|cccccccccc}
\hline\hline
$G^{(t)}(p,q,6)$ & p=1 & 2 & 3 & 4 & 5 & 6 & 7 & 8 & 9 & 10 \\
\hline
$q=0$ & $-\frac{4133r^{2}}{225}$ & $-\frac{524r^{4}}{27}$ &
$-\frac{21r^{6}}{5}$ & 0 & $\frac{7r^{10}}{360}$ & 
0 & 0 & 0 & 0 & 0 \\
1 & $-\frac{992}{99}$ & $-\frac{8266r^{2}}{225}$ &
$-\frac{524r^{4}}{27}$ & $-\frac{14r^{6}}{5}$ & 0 &
$\frac{7r^{10}}{900}$ & 0 & 0 & 0 & 0 \\
2 & $\frac{3834916}{96525r^{2}}$ & $-\frac{992}{33}$ &
$-\frac{4133r^{2}}{75}$ & $-\frac{524r^{4}}{27}$ &
$-\frac{21r^{6}}{10}$ & 0 & $\frac{7r^{10}}{1800}$ & 0 & 0 & 0 \\
3 & $\frac{233984}{1287r^{4}}$ & $\frac{15339664}{96525r^{2}}$ &
$-\frac{1984}{33}$ & $-\frac{16532r^{2}}{225}$ &
$-\frac{524r^{4}}{27}$ & $-\frac{42r^{6}}{25}$ & 0 &
$\frac{r^{10}}{450}$ & 0 & 0 \\
4 & $\frac{90142232}{546975r^{6}}$ & $\frac{1169920}{1287r^{4}}$ &
$\frac{7669832}{19305r^{2}}$ & $-\frac{9920}{99}$ &
$-\frac{4133r^{2}}{45}$ & $-\frac{524r^{4}}{27}$ &
$-\frac{7r^{6}}{5}$ & 0 & $\frac{r^{10}}{720}$ & 0 \\
5 & $-\frac{11427479552}{3741309r^{8}}$ &
$\frac{180284464}{182325r^{6}}$ & $\frac{1169920}{429r^{4}}$ &
$\frac{15339664}{19305r^{2}}$ & $-\frac{4960}{33}$ &
$-\frac{8266r^{2}}{75}$ & $-\frac{524r^{4}}{27}$ &
$-\frac{6r^{6}}{5}$ & 0 & $\frac{r^{10}}{1080}$ \\
\hline\hline
\end{tabular}

\vspace{40pt}
\begin{tabular}{c|cccccccccc}
\hline\hline
$G^{(v)}(p,q,6)$ & p=1 & 2 & 3 & 4 & 5 & 6 & 7 & 8 & 9 & 10 \\
\hline
$q=0$ & $-\frac{487r^{2}}{630}$ & $-\frac{1823r^{4}}{1890}$ &
0 & $\frac{r^{8}}{12}$ & $\frac{r^{10}}{144}$ & 
0 & 0 & 0 & 0 & 0 \\
1 & $\frac{3671}{3465}$ & $-\frac{487r^{2}}{315}$ &
$-\frac{1823r^{4}}{1890}$ & 0 & $\frac{r^{8}}{24}$ &
$\frac{r^{10}}{360}$ & 0 & 0 & 0 & 0 \\
2 & $\frac{1058444}{135135r^{2}}$ & $\frac{3671}{1155}$ &
$-\frac{487r^{2}}{210}$ & $-\frac{1823r^{4}}{1890}$ &
0 & $\frac{r^{8}}{40}$ & $\frac{r^{10}}{720}$ & 0 & 0 & 0 \\
3 & $\frac{12824}{715r^{4}}$ & $\frac{4233776}{135135r^{2}}$ &
$\frac{7342}{1155}$ & $-\frac{974r^{2}}{315}$ &
$-\frac{1823r^{4}}{1890}$ & 0 & $\frac{r^{8}}{60}$ &
$\frac{r^{10}}{1260}$ & 0 & 0 \\
4 & $-\frac{37335908}{765765r^{6}}$ & $\frac{12824}{143r^{4}}$ &
$\frac{2116888}{27027r^{2}}$ & $\frac{7342}{693}$ &
$-\frac{487r^{2}}{126}$ & $-\frac{1823r^{4}}{1890}$ &
0 & $\frac{r^{8}}{84}$ & $\frac{r^{10}}{2016}$ & 0 \\
5 & $-\frac{104536852792}{130945815r^{8}}$ &
$-\frac{74671816}{255255r^{6}}$ & $\frac{38472}{143r^{4}}$ &
$\frac{4233776}{27027r^{2}}$ & $\frac{3671}{231}$ &
$-\frac{487r^{2}}{105}$ & $-\frac{1823r^{4}}{1890}$ &
0 & $\frac{r^{8}}{112}$ & $\frac{r^{10}}{3024}$ \\
\hline\hline
\end{tabular}

\vspace{40pt}
\begin{tabular}{c|cccccccccc}
\hline\hline
$G^{(s)}(p,q,6)$ & p=1 & 2 & 3 & 4 & 5 & 6 & 7 & 8 & 9 & 10 \\
\hline
$q=0$ & $\frac{833r^{2}}{1350}$ & $\frac{1139r^{4}}{1890}$ &
$\frac{r^{6}}{5}$ & $\frac{r^{8}}{36}$ & $\frac{r^{10}}{720}$ & 
0 & 0 & 0 & 0 & 0 \\
1 & $\frac{137}{165}$ & $\frac{833r^{2}}{675}$ &
$\frac{1139r^{4}}{1890}$ & $\frac{2r^{6}}{15}$ &  
$\frac{r^{8}}{72}$ &
$\frac{r^{10}}{1800}$ & 0 & 0 & 0 & 0 \\
2 & $\frac{485768}{675675r^{2}}$ & $\frac{137}{55}$ &
$\frac{833r^{2}}{450}$ & $\frac{1139r^{4}}{1890}$ &
$\frac{r^{6}}{10}$ & $\frac{r^{8}}{120}$ & 
$\frac{r^{10}}{3600}$ & 0 & 0 & 0 \\
3 & $-\frac{362008}{135135r^{4}}$ & $\frac{1943072}{675675r^{2}}$ &
$\frac{274}{55}$ & $\frac{1666r^{2}}{675}$ &
$\frac{1139r^{4}}{1890}$ & $\frac{2r^{6}}{25}$ & 
$\frac{r^{8}}{180}$ & $\frac{r^{10}}{6300}$ & 0 & 0\\
4 & $-\frac{92273924}{3828825r^{6}}$ & $-\frac{362008}{27027r^{4}}$ &
$\frac{971536}{135135r^{2}}$ & $\frac{274}{33}$ &
$\frac{833r^{2}}{270}$ & $\frac{1139r^{4}}{1890}$ &
$\frac{r^{6}}{15}$ & $\frac{r^{8}}{252}$ & 
$\frac{r^{10}}{10080}$ & 0 \\
5 & $-\frac{15799818344}{130945815r^{8}}$ &
$-\frac{184547848}{1276275r^{6}}$ & $-\frac{362008}{9009r^{4}}$ &
$\frac{1943072}{135135r^{2}}$ & $\frac{137}{11}$ &
$\frac{833r^{2}}{225}$ & $\frac{1139r^{4}}{1890}$ &
$\frac{2r^{6}}{35}$ & $\frac{r^{8}}{336}$ & 
$\frac{r^{10}}{15120}$ \\
\hline\hline
\end{tabular}

\end{document}